\def\etal{{et~al.}}
\documentclass[10pt,preprint]{aastex}
\usepackage{subfigure}
\begin{document}
\setcounter{figure}{0}
\title{Proper Motions of Dwarf Spheroidal
Galaxies from \textit{Hubble Space Telescope} Imaging.  III:
Measurement for Ursa Minor.\footnote{Based on observations with NASA/ESA
\textit{Hubble Space Telescope}, obtained at the Space Telescope
Science Institute, which is operated by the Association of
Universities for Research in Astronomy, Inc., under NASA contract NAS
5-26555.}}

\author{Slawomir Piatek} \affil{Dept. of Physics, New Jersey Institute
of Technology,
Newark, NJ 07102 \\ E-mail address: piatek@physics.rutgers.edu}

\author{Carlton Pryor}
\affil{Dept. of Physics and Astronomy, Rutgers, the State University
of New Jersey, 136~Frelinghuysen Rd., Piscataway, NJ 08854--8019 \\
E-mail address: pryor@physics.rutgers.edu}

\author{Paul Bristow}
\affil{Space Telescope European Co-ordinating Facility, 
Karl-Schwarzschild-Str. 2, D-85748, Garching bei Munchen, Germany \\
E-mail address: bristowp@eso.org }

\author{Edward W.\ Olszewski}
\affil{Steward Observatory, University of Arizona,
    Tucson, AZ 85721 \\ E-mail address: eolszewski@as.arizona.edu}

\author{Hugh C.\ Harris}
\affil{US Naval Observatory, Flagstaff Station, P. O. Box 1149,
Flagstaff, AZ 86002-1149 \\ E-mail address: hch@nofs.navy.mil}

\author{Mario Mateo} \affil{Dept. of Astronomy, University of
Michigan, 830 Denninson Building, Ann Arbor, MI 48109-1090 \\
E-mail address: mateo@astro.lsa.umich.edu}

\author{Dante Minniti}
\affil{Universidad Catolica de Chile, Department of Astronomy and
Astrophysics, Casilla 306, Santiago 22, Chile \\
E-mail address: dante@astro.puc.cl}

\author{Christopher G.\ Tinney}
\affil{Anglo-Australian Observatory, PO Box 296, Epping, 1710,
Australia \\ E-mail address: cgt@aaoepp.aao.gov.au}

\begin{abstract}

	This article presents a measurement of the proper motion of the
Ursa Minor dwarf spheroidal galaxy determined from images taken with the
Hubble Space Telescope in two distinct fields.  Each field contains a
quasi-stellar object which serves as the ``reference point.''

	The measured proper motion for Ursa Minor, expressed in the
equatorial coordinate system, is $(\mu_{\alpha},\mu_{\delta})=(-50 \pm
17, 22 \pm 16)$~mas~century$^{-1}$.  Removing the contributions of the
solar motion and the motion of the Local Standard of Rest yields the
proper motion in the Galactic rest frame:
$(\mu_{\alpha}^{\mbox{\tiny{Grf}}}, \mu_{\delta}^{\mbox{\tiny{Grf}}}) =
(-8 \pm 17 ,38 \pm 16)$~mas~century$^{-1}$.  The implied space velocity
with respect to the Galactic center has a radial component of
$V_{r}=-75 \pm 44$~km~s$^{-1}$ and a tangential component of $V_{t}=144
\pm 50$~km~s$^{-1}$.

	Integrating the motion of Ursa Minor in a realistic potential
for the Milky Way produces orbital elements.  The perigalacticon and
apogalacticon are 40~$(10,76)$~kpc and 89~$(78,160)$~kpc,
respectively, where the values in the parentheses represent the $95\%$
confidence interval derived from Monte Carlo experiments.  The
eccentricity of the orbit is $0.39~(0.09,0.79)$ and the orbital period
is $1.5~(1.1,2.7)$~Gyr.  The orbit is retrograde and inclined by
$124~(94,136)$~degrees to the Galactic plane.

	Ursa Minor is not a likely member of a proposed stream of
galaxies on similar orbits around the Milky Way.  Nor is the plane of
its orbit coincident with a recently-proposed planar alignment of
galaxies around the Milky Way.  Comparing the orbits of Ursa Minor and
Carina shows no reason for the different star formation histories of
these two galaxies.  Ursa Minor must contain dark matter to have a high
probability of surviving disruption by the Galactic tidal force until
the present.

\end{abstract}

\keywords{galaxies: dwarf spheroidal --- galaxies: individual (Ursa Minor) ---
astrometry: proper motion}

\section{Introduction}
\label{intro}

	The Ursa Minor dwarf spheroidal galaxy (dSph), detected by
Wilson (1955) on plates of the Palomar Sky Survey, is at
$(\alpha,\delta)=(15^{\mbox{h}}09^{\mbox{m}}11^{\mbox{s}},67^{\circ}
12^{\prime}54^{\prime\prime})$ (J2000.0; Mateo 1998) on the sky.  Its
Galactic coordinates are $(l,b)=(104.95^{\circ}, 44.80^{\circ})$.
Several published estimates of the distance to Ursa Minor disagree by
more than their combined uncertainties.  Olszewski \& Aaronson (1985)
and Cudworth, Olszewski, \& Schommer (1986) determine a distance
modulus, $(m-M)_{0}$, of $19.0\pm 0.1$, which corresponds to a
heliocentric distance of $63\pm3$~kpc.  These two studies are not
independent because they use the same photometric calibration.  This
distance is in agreement with those measured by Nemec, Wehlau, \&
Mendes de Oliveira (1988),
$70 \pm 9$~kpc, and Mighell~\&~Burke (1999), $69 \pm 4$~kpc.  The
latter value comes from $V$ and $I$ images of the central region in
Ursa Minor taken with the {\it Hubble Space Telescope} (HST).
However, Carrera \etal\ (2002) determine $(m-M)_{0}=19.4 \pm 0.1$,
which gives a distance of $76 \pm 4$~kpc (note an erroneous entry of
70~kpc for the distance in their Table~3).  Bellazzini \etal\ (2002)
find a similar value: $(m-M)_{0}=19.41 \pm 0.12$.  About 0.3~mag of
the 0.41~mag difference between the distance moduli of Cudworth,
Olszewski, \& Schommer (1986) and Bellazzini \etal\ (2002) comes from
different values for the absolute magnitude of the horizontal branch.
It is beyond the scope of this article to resolve the uncertainties in
the globular cluster distance scale, however the larger distance is
based on more modern stellar models and is probably in better agreement
with Hipparcos parallaxes for subdwarfs (Reid 2003; Gratton
\etal\ 2003) and the age of the universe derived by WMAP (Bennett
\etal\ 2003).  Thus, this study adopts a distance of 76~kpc for the
purpose of deriving distance-dependent quantities.

	Ursa Minor is one of the least luminous Galactic dSphs.  Irwin
\& Hatzidimitriou (1995: IH hereafter) report its luminosity to be
$L_{V}=(2.0 \pm
0.9) \times 10^{5}~L_{\odot}$.  Reflecting their larger estimate of
the distance, Carrera \etal\ (2002) report $L_{V}=3 \times
10^{5}~L_{\odot}$.  These luminosities are based on the structural
parameters from IH: a core radius of $15.8
\pm 1.2$~arcmin and a tidal radius of $50.6 \pm 3.6$~arcmin.  Kleyna
\etal\ (1998) find similar structural parameters for Ursa Minor.
However, Mart\'{\i}nez-Delgado \etal\ (2001) and Palma \etal\ (2003)
find a more extended radial profile, which implies a larger
luminosity.  Palma \etal\ (2003) determine that both blue horizontal
branch stars and stars identified as giants on the basis of photometry
in the Washington-band system extend beyond the tidal radius of IH.
They argue that the luminosity of Ursa
Minor is 2.7$\times$ larger than the value of IH.  Based on both the larger radial extent and the larger
distance, Bellazzini \etal\ (2002) adopt a luminosity of $1.1 \pm 0.4
\times 10^6~L_{\odot}$.  Our study likewise adopts this value.

	Several studies find that the stars of Ursa Minor are old,
metal-poor, and of similar age.  Olszewski \& Aaronson (1985)
constructed a $V$ {\it versus} $B-V$ color-magnitude diagram for the
dSph using data from ground-based observations.  The diagram reveals
that the stars of Ursa Minor have ages and metallicities similar to
those of the metal-poor globular cluster M92.  The study interprets
the few stars brighter and bluer than the main-sequence turnoff as
blue stragglers.  Carrera \etal\ (2002) confirm this basic picture
using ground-based data in the $B$, $V$, $R$, and $I$ bands.  An
analysis of HST data by Mighell \& Burke (1999) reveals the same
picture: the stars of Ursa Minor are about 14~Gyr old, formed in a
single burst that lasted less than 2 Gyr, and are metal-poor --- with
a metallicity $\mbox{[Fe/H]} \approx -2.2$~dex.  All of these
characteristics indicate that the stars in Ursa Minor formed in a
single burst during the earliest stages of the formation of our
Galaxy.  Ursa Minor has little or no gas: searches for neutral
hydrogen by Young (2000) and for ionized hydrogen by Gallagher \etal\
(2003) yield no positive detections, only placing upper limits of
$7\times 10^3\ M_\odot$ on HI and $1\times 10^5\ M_\odot$ on HII.
The absence of gas is consistent with the old
stellar population in Ursa Minor; however, the question of why
Ursa Minor quickly stopped forming stars whereas the similar dSph
Carina did not remains unanswered.

	With the possible exception of Sagittarius, Ursa Minor has the
greatest flattening among the known Galactic dSphs.
IH derive an ellipticity of $0.56 \pm
0.05$ and a position angle for the major axis of $53 \pm 5$~degrees.
Kleyna \etal\ (1998) derive similar values.  The ellipticity of Ursa
Minor is distinctly larger than those of the other dSphs (again
excluding Sagittarius), which range from 0.1 to 0.35
(IH).  The left-hand panel in
Figure~\ref{fields} shows a $1^{\circ} \times 1^{\circ}$ section of
the sky in the direction of Ursa Minor from the Digitized Sky Survey.
The dashed and solid ellipses delineate the core and tidal radii of
Ursa Minor from IH.
	
	Hartwick \& Sargent (1978) and Aaronson (1983) were the first
to measure radial velocities for 1--2 stars in Ursa Minor, finding a
systemic velocity of about $-250$~km~s$^{-1}$.  Hargreaves \etal\
(1994) find a mean heliocentric radial velocity of $-249.2 \pm
1.5$~km~s$^{-1}$ from the radial velocities of 35 giants.  Armandroff,
Olszewski, \& Pryor (1995) combined the data from Hargreaves \etal\
(1994) and from Olszewski, Aaronson, \& Hill (1995) with their own
measurements to obtain mean radial velocities for a sample of 94
stars.  Excluding one possible non-member, the mean heliocentric
radial velocity is $-247.4 \pm 1.0$~km~s$^{-1}$.  Our study adopts
this last value.  The adopted value is in a reasonable agreement with
the most recent determination of this quantity by Wilkinson \etal\
(2004), who derive $-245.2^{+1.0}_{-0.6}$ km~s$^{-1}$ from the
measured radial velocities for 162 stars.

	The dispersion of velocities within a dSph provides information
about its mass and how the mass is distributed.  Aaronson \& Olszewski
(1987) report a velocity dispersion for Ursa Minor of $11 \pm
3$~km~s$^{-1}$ from radial velocities of seven stars.  This dispersion
indicated for the first time that the mass greatly exceeds that
expected from stars alone.  Subsequent measurements have confirmed the
large dispersion: Hargreaves \etal\ (1994) find
$7.5^{+1.0}_{-0.9}$~km~s$^{-1}$ from a sample of 35 giants and
Armandroff, Olszewski, \& Pryor (1995) find $8.8 \pm 0.8$~km~s$^{-1}$
from a sample of 93 giants.  The larger sample of radial velocities in
Wilkinson \etal\ (2004) provides information on the dependence of
velocity dispersion on the projected radius.  The bottom panel of their
Figure~1 shows that the velocity dispersion is constant at about
12~km~s$^{-1}$ to a radius of approximately 32~arcmin; beyond this
radius, the velocity dispersion drops sharply to about 2~km~s$^{-1}$.
Hargreaves \etal\ (1994) and Armandroff, Olszewski, \& Pryor (1995)
assume virial equilibrium, that light follows mass, and the luminosity
and structural parameters from IH to calculate a mass-to-light ratio in
solar units, $M/L_{V}$, of $50^{+36}_{-22}$ and $55 \pm 10$,
respectively.  The uncertainty for the second of these values includes
only the contribution from the velocity dispersion, whereas the
uncertainty for the first value includes the contributions from all of
the relevant parameters.  Making the same assumptions and using one of
the above velocity dispersions, IH, Kleyna \etal\ (1998), and
Carrera~\etal\ (2002) infer similarly large values of $M/L_V$: $95 \pm
43$, $70^{+30}_{-20}$, and 70, respectively.

	In contrast, Palma \etal\ (2003) find $M/L_{V} = 47$.  They
derive a surface density profile from a sample of stars selected to be
members of Ursa Minor using three-color Washington-band photometry.
Fitting a King model to this profile produces a tidal radius of $77.9
\pm 8.9$~arcmin.  This value is larger than those derived by
IH and Kleyna \etal\ (1998), which affects
the derived mass and luminosity of the galaxy.  Palma \etal\ (2003)
also note that the $M/L$ could be even smaller if the velocity
dispersion tensor of the dSph is anisotropic.  Richstone~\&~Tremaine
(1986) caution that the $M/L$ calculated using King's method will be
overestimated or underestimated depending on whether the velocity
dispersion is larger in the radial or tangential directions,
respectively.  Despite a lack of any evidence for an anisotropic
dispersion tensor, Palma \etal\ (2003) conclude that the $M/L_V$ of
Ursa Minor may be as low as 16.  The larger distance, hence larger
luminosity, found by Bellazzini~\etal\ (2002) further reduces the
lower limit on the $M/L_V$ to 7.

	Palma \etal\ (2003) note that a power law with an index of $-3$
is a better fit to the surface density profile of Ursa Minor at radii
between 20~arcmin and 100~arcmin than is a King model.  They argue that
such a profile indicates that the Galactic tidal force has produced a
halo of escaping stars around the dSph.  They also argue that a $M/L$
near the lower limit allowed by the uncertainty in the mass estimate is
more consistent with a picture in which ``this system is very likely
undergoing significant mass loss due to its tidal interaction with the
Milky Way.''

	G\'{o}mez-Flechoso~\&~Mart\'{\i}nez-Delgado (2003) have made the above
argument more quantitative.  They derive $M/L_V = 12$ for Ursa Minor
by evolving an $N$-body model of the dSph in a potential of the Milky
Way on an orbit constrained by the proper motion of the dSph from
Schweitzer, Cudworth, \& Majewski (1997), varying the $M/L$ of the
dSph in order to match the surface brightness profile of the model
dSph with the observed profile.  The study finds that an acceptable
match exists only if Ursa Minor has $M/L_V$ within a factor of two of
$12$.  G\'{o}mez-Flechoso \& Mart\'{\i}nez-Delgado (2003) argue that their
lower value for the $M/L$ is consistent with the values derived from
the velocity dispersion if the maximum possible effect from radial
anisotropy is taken into account.  They also argue that the observed
velocity dispersion ``could be inflated by the effects of the
substructures in the main body'' or by departures from virial
equilibrium due to the Galactic tidal force.

	The numerical simulations of Piatek \& Pryor (1995) show that,
if the Galactic tidal force causes a departure from virial
equilibrium, then escaping stars produce a velocity gradient --- an
apparent rotation --- along the projected major axis of the dSph.
Hargreaves \etal\ (1994) and Armandroff, Olszewski, \& Pryor (1995)
both found a statistically significant velocity gradient, but along
the minor axis instead of the major axis.  Subtracting the velocity
gradient from the radial velocities does not reduce the measured
velocity dispersion significantly.  Kroupa (1997) and Klessen \&
Kroupa (1998) show that the velocity gradient is hidden if the tidal
debris is aligned along the line of sight to the dSph.  Palma \etal\
(2003) rule out this alignment because it would produce more
broadening of the horizontal branch than they observe.  Thus, a
departure from virial equilibrium cannot explain the difference
between the $M/L$ found by G\'{o}mez-Flechoso \& Mart\'{\i}nez-Delgado (2003)
and those found using the measured velocity dispersion.

	Olszewski~\&~Aaronson (1985) discovered a statistically
significant change in surface density across their
3~arcmin~$\times$~5~arcmin field and interpreted this variation as
evidence for substructure in Ursa Minor.  It is now known that the
highest peak in the surface density of the dSph is near the eastern
edge of this field.  Demers \etal\ (1995) also detect a statistically
significant clump of stars at the same location, which causes a step in
the surface density profile at a radius of about 1~arcmin.  The deep
\textit{HST} photometry of Battinelli~\&~Demers (1999) supports the
existence of this clump and may also suggest the presence of additional
structure within the clump.  Other studies of substructure have focused
on a possible secondary peak or ``shoulder'' in the surface density
profile, first noted by IH, that is about 15~arcmin northeast of the
primary peak.  Kleyna \etal\ (1998), Eskridge \& Schweitzer (2001), and
Palma \etal\ (2003) demonstrate that a ``shoulder'' in the profile is
statistically significant, though a secondary peak is not.  Kleyna
\etal\ (2003) used measured radial velocities for stars in Ursa Minor
to demonstrate the presence of a distinct sub-population with a
velocity dispersion of 0.5~km~s$^{-1}$ --- much smaller than the
8.8~km~s$^{-1}$ of the whole sample of velocities --- at the location
of the ``shoulder.''  Excluding the stars in this substructure from the
sample would increase the measured velocity dispersion of the dSph and,
thus, the inferred $M/L$.  Thus, the presence of substructure does not
reconcile the $M/L$ found by G\'{o}mez-Flechoso \&
Mart\'{\i}nez-Delgado (2003) with that derived from the velocity
dispersion.

	Two pictures of Ursa Minor emerge from the above discussion.
(1) Dark matter is the main component of the mass of Ursa Minor and
therefore it determines the structure and internal dynamics of the
dSph.  (2) Ursa Minor contains little or no dark matter and therefore
the Galactic tidal force has had and continues to have an important
effect on the structure and internal dynamics of the dSph.  Knowing
the proper motion of Ursa Minor can help to discriminate between these
two pictures by constraining the orbit and, thus, the strength of the
Galactic tidal force on Ursa Minor.  If the perigalacticon is small,
Ursa Minor would need to contain a large amount of dark matter to have
survived until now.  The presence of tidal signatures need not
indicate that the dSph is out of virial equilibrium or that it
contains only luminous matter.  Conversely, a large perigalacticon
would require a low $M/L$ for the dSph so that the Galactic tidal
force could have produced the observed signatures.

	A proper motion for Ursa Minor also tests the hypothesis that
it is a member of a ``stream'' of objects in the Galactic halo that
share a similar orbit.  Lynden-Bell~\&~Lynden-Bell (1995) propose that
Ursa Minor, the Large Magellanic Cloud (LMC), the Small Magellanic
Cloud (SMC), Draco, and possibly also Sculptor and Carina, form a
stream.  They propose that such a stream forms from the fragments of a
larger, tidally disrupted galaxy.  The theory of streams in the
Galactic halo is falsifiable because it predicts a proper motion for
each member of the stream.  There is some controversy regarding the
reality of this stream: Anguita, Loyola, \& Pedreros (2000) measured
the proper motion of the LMC and found that it is inconsistent with the
prediction of Lynden-Bell~\&~Lynden-Bell (1995), whereas the proper
motions measured by Jones, Klemola, \& Lin (1994) and Kroupa \& Bastian
(1997) are consistent.

	There are two, independent, published results for the proper
motion of Ursa Minor.  Scholz~\&~Irwin (1993) used Palomar Sky Survey
plates for the first and second epochs, which are separated by about 26
years, and plates taken with the Tautenburg Schmidt telescope for the
third epoch --- which provide a total time baseline of about 31~years.
They find a proper motion, ``as is on the sky,'' of $(50\pm 80, 120\pm 80)$~mas~cent$^{-1}$ (the last row in their Table~3, but using an
uncertainty 80 instead of 20~mas~cent$^{-1}$ for $\mu_{\delta}$, as
implied by the text).  Schweitzer (1996) and Schweitzer, Cudworth, \&
Majewski (1997) used a total of 39 photographic plates taken with the
Palomar 5-meter Hale telescope (first epoch) and the KPNO 4-meter
telescope (second and third epochs).  The time between the first- and
third-epoch images is about 42 years.  Schweitzer (1996) reports a
proper motion of $(5.6\pm 7.8, 7.4\pm 9.9)$~mas~cent$^{-1}$.  The proper
motions measured by these two studies agree within their
uncertainties.

	Here we present a third independent measurement of the proper
motion of Ursa Minor.  The measurement derives from images of two
distinct fields, each containing a known quasi-stellar object (QSO),
taken with \textit{HST}.  Sections~\ref{data}, \ref{analysis}, and
\ref{pm} describe the data, the derivation of mean centroids at each
epoch, and the derivation of the proper motion from the centroids,
respectively.  The last of these sections contains a comparison of our
measured proper motion with those determined by Scholz~\&~Irwin (1993)
and Schweitzer (1996, see also Schweitzer, Cudworth, \& Majewski
1997).  Section~\ref{orbitsec}\ derives orbital elements in a realistic
potential for the Milky Way.  Section~\ref{sec:disc} discusses the
implications of the orbit of Ursa Minor for:  1) its membership in
proposed structures in the Galactic halo; 2) its star formation
history; 3) the mass of the Milky Way; and 4) the importance of the
Galactic tidal force and, thus, whether the dSph contains dark matter.
Finally, Section~\ref{sec:summary} summarizes our main results.

\section{Observations and Data}
\label{data}

	The data comprise images of two distinct fields in the
direction of Ursa Minor taken at three epochs with {\it HST}.  A known
QSO is at, or close to, the center of each field.  The left panel in
Figure~\ref{fields} shows the locations of the two fields on the sky.
Both fields are within the core radius and close to the minor axis.
\textit{HST} imaged the larger of the two fields, UMI~J$1508+6716$, on
February 16, 2000; February 15, 2001; and February 1, 2002 using the
Space Telescope Imaging Spectrograph (STIS) with no filter (50CCD).
Ursa Minor was in the continuous viewing zone during the first two
epochs and was not during the last.  Each of the eight dither
positions has six images at the first two epochs, for a total of 48
images per epoch, while each dither position has only three, for a
total of 24 images, at the last epoch.  The average exposure times for
the three epochs are 176~s, 176~s, and 210~s, respectively.  The QSO
in this field is at $(\alpha,\delta)=(15^{\mbox{h}}08^{\mbox{m}}37\fs
661,+67^{\circ} 16^{\prime}34\farcs 27)$ (J2000.0).  From Bellazzini
\etal\ (2002), its $V$- and $I$-band magnitudes are 20.3 and 19.7,
respectively.  Figure~\ref{fig:QSO-STIS} shows a spectrum of the QSO,
which indicates a redshift of 1.216.  The top-right panel in
Figure~\ref{fields} shows a sample image of this field from the 2000
epoch.  The image is the average of six images at one dither
position with cosmic rays removed.  The cross-hair indicates the
location of the QSO.

	\textit{HST} imaged the smaller of the two fields,
UMI~J$1508+6717$, on March 14, 1999; March 10, 2001; and March 2, 2003
using the Planetary Camera (PC) of the Wide Field and Planetary Camera
2 (WFPC2) and the F606W filter.  There are 40 images for the first
epoch, 36 for the second, and 36 for the third.  The exposure time is
160~s for each image.  Here, the QSO is at
$(\alpha,\delta)=(15^{\mbox{h}}08^{\mbox{m}}40\fs 410,+67^{\circ}
17^{\prime}47\farcs 50)$ (J2000.0), its $V$- and $I$-band magnitudes
are $19.7$ and $18.9$, respectively (Bellazzini \etal\ 2002), and
Figure~\ref{fig:QSO-WFPC} shows its spectrum.  The spectrum implies a
redshift of 0.716.  The bottom-right panel of Figure~\ref{fields}\
depicts an average of three images in this field from the 1999 epoch,
with cosmic rays removed.  The cross-hair indicates the location of
the QSO.  Note the scarcity of stars in both sample images.

	Bristow (2004) notes that the decreasing charge transfer
efficiency in the STIS and WFPC2 CCDs may induce a spurious
contribution to the proper motion measured with the method described by
Piatek \etal\ (2002: P02 hereafter) and Piatek \etal\ (2003; P03
hereafter).  The cosmic particle radiation damages the crystal lattice
of a detector, creating ``charge traps,''  and the number of these
traps increases with time.  During the readout, charge moves along the
$Y$ axis: ``up,'' or towards the increasing $Y$-values, for STIS, and
``down,'' or in the direction of decreasing $Y$-values, for PC2.
Charge moves ``left,'' towards decreasing $X$-values in the serial
register.  When a ``packet'' of charge corresponding to some object
encounters a trap, it loses some of its charge.  A passing packet
partially fills a trap, so a subsequent packet loses less charge.
Streaks of light along the $X$ and $Y$ axes seen trailing behind
objects in the images are the visual artefacts caused by the gradual
release of charge from the traps.  The loss of charge for motion in the
$X$ direction is smaller than for that in the $Y$ direction; it is
negligible for STIS (Brown \etal\ 2002) and one third as large for
WFPC2 (Heyer \etal\ 2004).  The subsequent discussion focuses only on
the effect in the $Y$ direction.

The loss of charge to traps causes both the flux and
the PSF of an object to depend on its $Y$ coordinate in the image.  The
farther a packet travels, the more charge it loses.  Hence, the
measured flux of an object far from the serial register is smaller than
that for an identical object closer to the serial register; in other
words, the measured flux becomes position dependent.  This effect is
immaterial to our method.  However, as the packets representing an
object move along the $Y$ axis, those on its leading side fill
partially each trap encountered, so that there are fewer traps
available to remove charge from subsequent packets.  This non-uniform
loss of charge across the object changes its PSF.  Consequently, the
centroid of the packet shifts in the direction opposite to the readout
direction.  The shift is greatest for objects that are farthest from
the serial register.  It is also larger for faint objects than for
bright objects (Bristow \& Alexov 2002).  This effect changes the
measured proper motion only because the number of charge traps
increases with time, thus causing the shift of a centroid to also
change with time.  The analysis of Bristow (2004) warns that this
spurious proper motion may be comparable to the actual proper motion of
a dSph.

	There are two possible approaches to removing the effect of the
spurious shifts of the centroids along the $Y$ axis with time.  In the
first approach, the images are corrected by restoring the trapped
charge to its original packet.  This correction requires that the
distribution and the properties of the charge traps be known.  Bristow
\& Alexov (2002) constructed a physical model of the charge traps and
then wrote a computer code that makes the aforementioned correction for
an image taken with STIS.  Unless stated otherwise, the results
presented are for corrected images.  No comparable software exists for
WFPC2, unfortunately.  A second approach to correcting for the
decreasing charge transfer efficiency is including a term that varies
linearly with $Y$ in the equations which transform the coordinates of
an object at different epochs to a common coordinate system.  The
coefficients of these transformations are fitted as described in
Section~\ref{sec:dpm}.  The analysis of the data taken with WFPC2 uses
this approach.

\section{Analysis of Data}
\label{analysis}

	The analysis of the data and derivation of the proper motion of
Ursa Minor are basically the same as those described in P02.  P03
provides some additional insight into the method.  This article only
lists the principal steps here and comments --- where it is appropriate
--- on several changes in the details of the analysis.  The principal
steps of the analysis are: (1) Using the DAOPHOT and ALLSTAR software
packages (Stetson 1987, 1992, 1994), determine an initial estimate of
the centroid of each object --- stars and the QSO.  (2) Construct an
effective point-spread function (ePSF; Anderson \& King 2000) for each
epoch separately from a select set of stars and the QSO.  (3)
Iteratively, fit the ePSF to an image of an object using least-squares
to derive a more accurate centroid of the object and, using these more
accurate centroids, derive a more accurate ePSF.  Repeat until a stable
solution ensues.  There may be a maximum of $N$ centroids for an
object, where $N$ is the number of images.  (4) Transform the centroids
measured at one epoch to a common --- fiducial --- coordinate system
and calculate the average.  The fiducial coordinate system for an epoch
is the $X-Y$ coordinate system of the chronologically first image.  (5)
Determine the transformation between the fiducial coordinate systems of
the later epochs and that of the first epoch using stars which are
likely members of Ursa Minor and common to all three epochs.  The fit
for the transformation simultaneously determines the change with time
of the coordinates of some objects, $\mu_{x}$ and $\mu_{y}$ in
pixel~yr$^{-1}$.  Ideally, these objects are not members of Ursa Minor.
The set of objects whose motion is fit is built starting with the QSO
and iteratively adding one object at a time, in order of descending
$\chi^{2}$ for the scatter about the mean coordinate, until the highest
$\chi^{2}$ of an object not yet in the set is below a specified limit.
The shifts are with respect to stars of Ursa Minor which, by
definition, have $\mu_{x} = \mu_{y} = 0$.  This procedure for deriving
the values of $\mu_{x}$ and $\mu_{y}$ for objects is new to our method;
Section~\ref{pm} below discusses its details.  The advantage of this
new method over that used by P02 and P03 is that it more accurately
includes the contribution from the uncertainty in the coordinate
transformation between epochs to the uncertainty in the proper motion.
(6) Derive the proper motion of Ursa Minor from the $\mu_{x}$ and
$\mu_{y}$ of the QSO.

Before calculating the transformation to the fiducial coordinate
system, the centroids measured in each image taken with WFPC2 are
corrected for the 34th-row defect (Shaklan, Sharman, \& Pravdo 1995;
Anderson \& King 1999).  In addition, all centroids are corrected for
the known optical distortions in the WFPC2 and STIS instruments.  This
paper uses the most recent corrections for WFPC2 derived by Anderson \&
King (2003).  The corrections for STIS continue to be those in the STIS
Data Handbook (Brown \etal\ 2002).

	The following subsections present and discuss the key
diagnostics of the performance of our method.

\subsection{Flux Residuals}
\label{rf}

	Equation~22 in P02 defines a flux residual diagnostic, ${\cal
RF}$, which is a quantitative measure of how the shape of an object
matches the ePSF.  Ignoring random noise, for a perfect match, ${\cal
RF}=0$; otherwise, ${\cal RF} \ne 0$, where the sign depends on the
details of the mismatch and the size of the mismatch increases with
the brightness of the star.  A plot of ${\cal RF}$ as a function of
location can help unravel a dependence of true PSF with location in an
image.  Our method uses a single, location-independent ePSF for
deriving centroids of objects in all images for a given field and
epoch.  If the true PSF of an object varies from image to image, then
the ${\cal RF}$s for this object will show a scatter, for example
around ${\cal RF}=0$ if the ePSF represents some average of the true
PSFs.  The plot of ${\cal RF}$ as a function of location will show
trends if the true PSF varies with location in the image.

	Figure~\ref{rf-stis} shows plots of ${\cal RF}$ as a function
of location for the UMI~J$1508+6716$ field.  The left-hand panels show
plots of ${\cal RF}$ {\it versus} $X$ and the right-hand panels show
plots of ${\cal RF}$ {\it versus} $Y$.  From top to bottom, the rows
of plots are for epochs 2000, 2001, and 2002.  Each plot combines
points from all images for a given epoch.  The solid square marks the
points corresponding to the QSO.  In all six panels, the values of
${\cal RF}$ for the QSO are negative and distinct from those for the
other objects --- stars.  The panel~\ref{stis-10-rf-x} and, to a lesser
extent, panel~\ref{stis-9-rf-x} show a linear trend between ${\cal
RF}$ and $X$.  ${\cal RF}$ tends to be positive around $X
\approx 0$~pixel and negative around $X \approx 1000$~pixel.  No other
panels show a trend of ${\cal RF}$ with $X$ or $Y$.

	Figure~\ref{rf-stis} implies the following.  (1) The PSF of
the QSO has a different shape than that of a star.  The ${\cal RF}$s
for the QSO become comparable to those for the stars when the value of
the central pixel in the $5\times 5$ science data array for the QSO
(see P02 for a description of the structure of the data) is
arbitrarily reduced by about 20\%.  Thus, the PSF of the QSO is
narrower than that of the stars, perhaps because the QSO is bluer than
the stars.  (2) The PSF at a given location in the field varies from
image to image, \textit{i.e.}, with time.  This variation causes the
scatter in the ${\cal RF}$s of a given object.  (3) The PSF varies
across the field for epoch 2002 and, to a lesser degree, for epoch
2001 because the plots for those epochs show trends with location.

	Figure~\ref{rf-pc} is analogous to Figure~\ref{rf-stis} for the
UMI~J$1508+6717$ field.  From the top row to the bottom, the plots
correspond to epochs 1999, 2001, and 2003, respectively.  All of the
plots show that the ${\cal RF}$s for the QSO are negative, with most of
the points outside of the plot for epoch 2001.  The mean ${\cal RF}$ of
the QSO at any epoch is more negative than the mean for any other
object. For example, the object at $(X,Y)\approx(720,300)$ is a star as
bright as the QSO but its ${\cal RF}$s tend to be less negative or even
positive.  Again, the likely reason for the distinctness of the QSO is
its narrower PSF.  No plot shows an evident trend of ${\cal RF}$ with
$X$ or $Y$; however, the small number of objects contributing to these
plots makes this conclusion uncertain.

\subsection{Position Residuals}
\label{rxry}

	Figures~\ref{stis-rxry} and \ref{pc-rxry} plot the position
residual, ${\cal RX}$ or ${\cal RY}$, of a centroid as a function of
its location within a pixel --- pixel phase, $\Phi_{x}$ or $\Phi_{y}$.  The
position residuals are ${\cal RX}\equiv \langle
X_{0}\rangle-X_{0}$ and ${\cal RY}\equiv \langle Y_{0}\rangle-Y_{0}$,
respectively, where $\langle X_{0}\rangle$ and $\langle Y_{0}\rangle$
are the components of the mean centroid in the fiducial coordinate
system --- the system of the first image in time at a given epoch.  The
pixel phases are $\Phi_{x}\equiv X_{0}-Int(X_{0})$ and $\Phi_{y}\equiv
Y_{0}-Int(Y_{0})$, where the function $Int(x)$ returns the integer
part of a variable $x$.  In the presence of only random noise,
the points in Figures~\ref{stis-rxry} and \ref{pc-rxry} would scatter
symmetrically around zero and would not exhibit any trends.

	Panels \ref{stis-8-rxry}, \ref{stis-9-rxry}, and
\ref{stis-10-rxry} in Figure~\ref{stis-rxry} plot, from top to bottom,
${\cal RX}$ {\it versus} $\Phi_{x}$, ${\cal RY}$ {\it versus} $\Phi_{y}$,
${\cal RX}$ {\it versus} $\Phi_{y}$, and ${\cal RY}$ {\it versus}
$\Phi_{x}$.  The panels are for the UMI~J$1508+6716$ field, epochs
2000, 2001, and 2002, respectively.  No plot in any of the panels in
Figure \ref{stis-rxry} shows a trend of position residual with
pixel phase.

	Figure~\ref{pc-rxry} is analogous to
Figure~\ref{stis-rxry} for the UMI~J$1508+6717$ field.  Panels
\ref{pc-7-rxry}, \ref{pc-9-rxry}, and \ref{pc-10-rxry} correspond to
epochs 1999, 2001, and 2003, respectively.  No plot in
Figure~\ref{pc-rxry} shows a trend of position residual with
pixel phase.

	The figures show that the values of ${\cal RX}$ and ${\cal RY}$
for the QSO are indistinguishable from those for the stars.  The smaller
scatter of the values for the QSO reflects its status as one of the
brightest objects in the field.  The similar distributions of the points
for the QSO and the stars implies that the narrower PSF of the QSO,
compared to that for the stars, does not affect the accuracy of its
measured centroid.

\subsection{Systematic error in the centroid of an object}
\label{serror}

	Kuijken~\&~Rich (2002) show that the precision of a centroid
determined by PSF fitting is proportional to $(S/N)^{-1}$ times the
FWHM of the PSF.  The constant of proportionality is approximately
0.67.  If there is no source of error other than the uncertainty in the
intensity registered by the pixel, then a plot for the entire sample of
the $rms$ scatter of the measured centroids for an object around their
mean as a function of the $(S/N)^{-1}$ of the object should consist of
points falling on a straight line passing through the origin.  If the
points fall above this straight line at large $S/N$, they indicate the
presence of additive uncertainty, either random or systematic, that is
independent of the signal.  The distribution of points above and
below the line is affected by the sampling uncertainty in the $rms$ and
variations in the FWHM of the PSF.  The PSF can vary with location in
the field, from image to image --- thus with time, or both.

	Figure~\ref{stis-8-rms-sn} is a plot of the $rms$ of the
$X$-component of a centroid (top panel) and of the $Y$-component
(bottom panel) as a function of $(S/N)^{-1}$ for the epoch 2000
UMI~J$1508+6716$ field.  Figures~\ref{stis-9-rms-sn} and
\ref{stis-10-rms-sn} are the same for epochs 2001 and 2002,
respectively.  Note that the figures have different horizontal and
vertical scales.  The solid line in each
plot is the best-fitting function of the form
\begin{equation}
\sigma = ((a (S/N)^{-1})^2 + \sigma_0^2)^{1/2},
\label{rms-sn}
\end{equation}
where $\sigma$ is
the $rms$ scatter in either the $X$ or $Y$ directions and $a$ and
$\sigma_0$ are free parameters.  Table~1 tabulates the fitted values
of $a$ and $\sigma_0$.  Each point has equal weight in the fit.
In all of the plots, the adopted functional form is a good fit to the
points and the best fit requires a non-zero $\sigma_0$.  The
fitted slopes are in approximate agreement with the value expected
for the 1.5~pixel FWHM of our ePSF.  The points corresponding to
the QSO are not farther from the fitted line than those corresponding
to bright stars, which argues that the difference between the PSF of
the QSO and of a star does not affect the $rms$ scatter of the
measured centroid of the QSO.

	Figures~\ref{pc-7-rms-sn}, \ref{pc-9-rms-sn}, and
\ref{pc-10-rms-sn} are the corresponding plots of $rms$ {\it versus}
$(S/N)^{-1}$ for the UMI~J$1508+6717$ field epochs 1999, 2001, and
2003, respectively.  Note that the figures have different horizontal and
vertical scales.  The solid lines show that Equation~\ref{rms-sn}\ is
again a good fit to the points.  Table~1 contains the best-fitting $a$
and $\sigma_0$ for all three epochs.  The figures and best-fitting
values are similar to those for the other field.

	The additive uncertainty, $\sigma_0$, inferred from the fits
shown in Figures~\ref{stis-rms-sn} and \ref{pc-rms-sn} significantly
increases the uncertainty of the final average centroid for objects
with high $S/N$.  P02 note that the $rms$ scatter of measured centroids
about their mean is smaller for measurements at a single dither
position than for those at multiple dither positions.  The
interpretation of this was that the errors producing the additive
uncertainty depend primarily on pixel phase, which argues that they
arise from errors in the shape of the ePSF.  Thus, the measurements of
the centroid at a single dither position are not independent when the
$S/N$ is high.  Both P02 and P03 calculate the larger uncertainty in
the mean centroid resulting from the smaller number of independent
measurements by assuming that, when the $rms$ of a component of the
centroid approaches the corresponding $\sigma_{0}$, the uncertainty in
the average centroid approaches the $rms$ divided by the square-root of
the number of dither positions instead of the square-root of the number
of measurements (see Equation~29 in P02).  Additional testing while
analyzing data for this article shows that the method of P02 and P03
overestimates the uncertainties in the centroids for those objects with
high $S/N$, while underestimating them for those with low $S/N$.

The usual estimator for the standard deviation about the mean is biased
downwards when the sample size is small (Sclove 2004).  For example,
the bias is a factor of 0.89 for a sample size of three.  This partly
explains why P02 finds a smaller $rms$ scatter around the mean
for measured centroids
at a given dither position compared to the scatter for the centroids at
all of the dither positions together.  A better method to estimate
the uncertainty in the final average centroid than
that used by P02, including the effects
of a lack of independence of measurements at a given dither position,
is to use the mean centroid, $\bar{z}_k$, at each dither position $k$.
The uncertainty in either the $x$ or $y$ component of the final centroid
is
\begin{equation}
\sigma_z = \left(\frac{1}{N_d} \sum_{k=1}^{N_d}
(\bar{z}_k - \langle \bar{z} \rangle)^2\right)^{1/2},
\label{eq:uncnew}
\end{equation}
where $N_d$ is the number of dither positions and $\langle \bar{z} \rangle$
is the mean of the mean centroids at the dither positions.  The uncertainty
estimated with equation~\ref{eq:uncnew} is typically 20\% -- 40\% larger than
the usual estimate from all of the measured centroids for an object with a
$S/N$ larger than about 15.  This increase is less than the approximately
70\% increase implied by the procedure of P02, but the increase extends
to a lower $S/N$.  This article uses Equation~\ref{eq:uncnew} to calculate the uncertainty in the mean centroid of an object at one epoch. 

\section{The Proper Motion of Ursa Minor}
\label{pm}

	The procedure for deriving a proper motion in this article is
different from that described in P02 and P03.  Section~\ref{sec:dpm}
below describes this new procedure in detail for the three-epoch data
taken either with STIS or WFPC2.  The subsequent sections describe the
actual realization of this procedure for the case of Ursa Minor.

\subsection{Deriving the Motion of the QSO in the
Standard Coordinate System}
\label{sec:dpm}

The three elements in determining the relative motion of the QSO with
respect to the stars of the dSph are:  1) A transformation between the
fiducial coordinate systems of the later epochs and that of the first
epoch --- the ``standard coordinate system'' --- determined by objects
common to all three epochs.  The transformation contains a translation,
$(\delta x_{j1}, \delta y_{j1})$, a rotation, $\theta_{j1}$, and a
ratio between the two scales, $s_{j1}$.  Here $j$ is the index denoting
the epoch and, for a transformation, may be 2 or 3.  2)  A mean
position for an object $i$, $(\bar{x}^{i},\bar{y}^{i})$, which is the
average of the three measured centroids in the standard coordinate
system.  The transformed standard coordinates at epochs 2 and 3 are
related to the mean coordinate by
$(x_{j}^{\prime i}, y_{j}^{\prime i}) = (\bar{x}^{i}, \bar{y}^{i}) +
(\mu_{x}^{i},\mu_{y}^{i})\, t(j)$
for the QSO and for any object with a large $\chi^2$ for the scatter
around $(\bar{x}^{i},\bar{y}^{i})$ with $(\mu_{x}^{i},\mu_{y}^{i}) =
(0, 0)$.  Here $(\mu_{x}^{i},\mu_{y}^{i})$ is the uniform linear motion
of the object in the standard coordinate system in pixel~yr$^{-1}$ and
$t(j)$ is the time of the epoch $j$ measured from $t(1)\equiv 0$.  3)
A simultaneous fit for the coefficients of the transformations, the
$(\bar{x}^{i},\bar{y}^{i})$, and the $(\mu_{x}^{i},\mu_{y}^{i})$ using
all of the measured centroids of objects common to the three epochs.
Simultaneously fitting for the $(\mu_{x}^{i},\mu_{y}^{i})$ ensures that
the stars of the dSph remain at rest in the standard coordinate
system.

	Let $(x_{j}^{i}\pm\sigma_{xj}^{i},y_{j}^{i}\pm\sigma_{yj}^{i})$
be the measured coordinates and their uncertainties of the centroid of
object $i$ in the fiducial coordinate system of epoch $j$, where $j=1$,
2, or 3.  The transformation of these measured coordinates to the
standard coordinate system is
\begin{eqnarray}
x_{j}^{\prime\, i} &=& x_{off} + \delta x_{j1} + s_{j1}\left((x_{j}^{i} -
x_{off})\cos\theta_{j1} - (y_{j}^{*i} - y_{off})\sin\theta_{j1} \right)
\label{eq:tranx} \\ 
y_{j}^{\prime\, i} &=& y_{off} + \delta y_{j1} + s_{j1}\left((x_{j}^{i} -
x_{off})\sin \theta_{j1} + (y_{j}^{*i} - y_{off})\cos \theta_{j1} \right)
\label{eq:trany} \\
\sigma_{xj}^{\prime\, i}&=&
s_{j1}\sqrt{(\sigma_{xj}^{i})^{2}\cos^{2}\theta_{j1}
+ (\sigma_{yj}^{i})^{2}\sin^{2}\theta_{j1}} \\
\sigma_{yj}^{\prime\, i}&=&
s_{j1}\sqrt{(\sigma_{xj}^{i})^{2}\sin^{2}\theta_{j1}
+ (\sigma_{yj}^{i})^{2}\cos^{2}\theta_{j1}}.
\label{eq:tranuy}
\end{eqnarray}
The offset $(x_{off}, y_{off})$ defines the pivot point for the
rotation: it is $(512, 512)$~pixel for STIS and $(400, 400)$~pixel for
WFPC2.  The $Y$ coordinate in equations~\ref{eq:tranx}\ and
\ref{eq:trany}, $y_{j}^{*i}$, includes a correction for the shift
caused by the charge traps in the CCD.  As discussed at the end of
Section~\ref{data}, the equations
\begin{eqnarray}
y_{j}^{*i} & = &  y_{j}^{i} +
b \frac{t(j)}{t(3)}(1024-y_{j}^{i}) \label{eq:cti-stis} \\
y_{j}^{*i} & = &  y_{j}^{i} +
b \frac{t(j)}{t(3)}y_{j}^{i}. \label{eq:cti-pc}
\end{eqnarray}
approximately correct for the shift in the $y$ direction induced by the
charge traps.  Equation~\ref{eq:cti-stis} is for the data taken with
STIS, whereas Equation~\ref{eq:cti-pc} is for those taken with WFPC2.
These equations are necessary only when the method of Bristow \& Alexov
(2002) has not been used to restore the images.  Because charge traps
affect a faint object more than a bright object, an object
contributes to the determination of the fitted parameter $b$, and has
its coordinates subsequently corrected, only if its $S/N$ is smaller
than some specified limit.

For the three-epoch data, equations~\ref{eq:tranx}\ -- \ref{eq:tranuy}
and \ref{eq:cti-stis}\ or \ref{eq:cti-pc}\ contain nine fitted parameters.
Their values result from
minimizing a $\chi^{2}$ of the form
\begin{equation}
\label{eq:k2}
\chi^{2}=\sum_{j=1}^{3} \sum_{i=1}^{N} \left[\left(\frac{x_{j}^{i} -
\left(\bar{x}^{i}+\mu_{x}^{i}t(j)\right)}{\sigma_{xj}^{\prime\,
i}}\right)^2 + \left(\frac{y_{j}^{i} -
\left(\bar{y}^{i}+\mu_{y}^{i}t(j)\right)}{\sigma_{yj}^{\prime\,
i}}\right)^2\right].
\end{equation}
The $(\bar{x}^{i},\bar{y}^{i})$ in the above equations add an
additional $2\times N$ fitted parameters, where $N$ is the number of
objects in the fit.  However, the $(\bar{x}^{i},\bar{y}^{i})$ can be
calculated analytically.  The minimization procedure starts by fitting
for the $(\mu_{x}^{i},\mu_{y}^{i})$ of the QSO while assigning
$(\mu_{x}^{i},\mu_{y}^{i})=(0,0)$~pixel~yr$^{-1}$ for all of the other
objects.  The procedure then iteratively selects the object with the
largest contribution to the $\chi^{2}$ and fits for its
$(\mu_{x}^{i},\mu_{y}^{i})$ together with all of the previously fit
parameters.  The iteration terminates when the highest $\chi^{2}$ among
the objects not yet selected is smaller than some specified limit.  The
proper motion of a dSph derives from $(-\mu_{x},-\mu_{y})$ for the
QSO.

	The uncertainty in $(\mu_{x},\mu_{y})$ for the QSO and, thus,
the uncertainty in the proper motion of the dSph comes from increasing
the $\chi^{2}$ by one above the minimum (\textit{e.g.}, Press
\etal\ 1992, Ch.~15).  One component of the motion changes away from
its fitted value, with all other parameters adjusted to minimize
$\chi^{2}$, until the total $\chi^{2}$ increases by one.  The
difference between this value and the nominal value is the uncertainty
for this component of the motion.  The procedure repeats for the other
component of the uncertainty.  A correct determination of the
uncertainty in $(\mu_{x},\mu_{y})$ using this method requires that the
uncertainties in the $(x_{j}^{i}, y_{j}^{i})$ be realistic.  A
break-down of this assumption would be indicated by a minimum total
$\chi^2$ that is significantly larger than one per degree of freedom.

\subsection{Motion of the QSO in the UMI~J$1508+6716$ field}
\label{sec:stis-qso}

	STIS was the imaging detector in the UMI~J$1508+6716$ field at
all of the epochs.  Thus, it was possible to correct the images for the
effects induced by the charge traps in the CCD using the method of
Bristow \& Alexov (2002).  Correcting the images eliminates the need to
fit for the free parameter $b$ in Equation~\ref{eq:cti-stis}.

	The number of objects with measured centroids is 81 for the
first epoch, 49 for the second, and 64 for the third.  Among these, 32
are common to the three epochs.  The choice for the individual $\chi^2$
that triggers fitting for uniform linear motion is 15.  There are
approximately 4 degrees of freedom per object when the motion is not
fit, so this limit should be triggered by chance for only 0.15 object
in a sample of 32.  The final fitted transformation and motion of the
QSO are not sensitive to the exact value of this limit.  The total
$\chi^2$ from equation~\ref{eq:k2} is much larger than one per degree
of freedom when using the uncertainties estimated with
equation~\ref{eq:uncnew}.  Thus, there must be an uncertainty that
arises from an error that is the same for all measurements made at a
single epoch, but changes significantly between epochs.
Figure~\ref{fig:k2stis} explores whether this uncertainty is additive
or multiplicative by plotting the individual contribution of each
object to the $\chi^2$ \textit{versus} $S/N$.  From top to bottom, the
panels show the contribution to $\chi^2$ for measurements at the first
epoch only, second only, third only, and at all of the epochs.  The
plots show that $\chi^2$ is, on average, the same at all values of
$S/N$.  Increasing the uncertainties given by equation~\ref{eq:k2} with
an additional additive uncertainty would decrease the $\chi^2$ values
only at high $S/N$, so Figure~\ref{fig:k2stis} indicates that the
uncertainty given by equation~\ref{eq:uncnew} should be multiplied by a
constant instead of having a constant added in quadrature.  Multiplying
the uncertainties by 1.55 produces a $\chi^2$ of one per degree of
freedom.  This choice for the multiplicative factor ensures that the
uncertainties in the proper motions calculated as described at the end
of Section~\ref{sec:dpm} reflect the true scatter of the points about
the fitted transformation and proper motions.  The $\chi^2$ values in
Figure~\ref{fig:k2stis} are calculated with the increased uncertainty.
We can point to no likely source for the additional error operating
between epochs.

	Figure~\ref{fig:stis-rxry} plots position residuals, defined
for an object $i$ by
\begin{eqnarray}
RX_{j-1}^{i}&=&\bar{x}^{i} + \mu_{x}^{i}t(j) - x_{j}^{\prime\, i}
\label{eq:rx} \\
RY_{j-1}^{i}&=&\bar{y}^{i} + \mu_{y}^{i}t(j) - y_{j}^{\prime\, i},
\label{eq:ry}
\end{eqnarray}
as a function of location in the standard coordinate system.  Here the
subscript $j-1$ indicates that the residual is for a centroid from the
$j^{th}$ epoch transformed to the standard coordinates system of the
time of the first epoch.  From top
to bottom, the panels are for the first, second, and third epoch,
respectively.  The panels in the left column show $RX$ \textit{versus}
$X$ and those in the right column show $RY$ \textit{versus} $Y$.  In
all of the plots, the points scatter around the horizontal axis;
no plot shows a trend between $RX$ and $X$ or $RY$ and $Y$ or a
systematic bias towards either positive or negative residuals.  The
scatter is a few hundredths of a pixel.  Although not
shown in the figure, the plots of $RX$ \textit{versus} $Y$ and of $RY$
\textit{versus} $X$ do not show any trends either.

	Figure~\ref{xqyqstis} shows the location of the QSO as a
function of time in the standard coordinate system.  The top panel
shows the variation of the $X$ coordinate and the bottom panel does
the same for the $Y$ coordinate.  The motion of the QSO is
$(\mu_{x},\mu_{y})=(0.0137\pm 0.0044,-0.0008\pm 0.0042)$~pixel~yr$^{-1}$.
The contribution to the total $\chi^2$ from the QSO is 0.17, which is
reflected in the close agreement between the points and the straight
lines in Figure~\ref{xqyqstis}.  The contribution to the $\chi^2$ has
approximately two degrees of freedom, which implies an 8\% probability
of a $\chi^2$ smaller than 0.17 by chance.

\subsection{Motion of the QSO in the UMI~J$1508+6717$ field}
\label{sec:pc-qso}

	WFPC2 was the imaging detector for this field.  Because no
corresponding software exists that restores an image taken with this
detector, equation~\ref{eq:cti-pc} must be used to account for the
shifts in the $Y$ direction caused by the charge traps.  The number of
objects with measured centroids is 21 for the 1999 epoch, 20 for 2001,
and 19 for 2003.  Among these, 19 are common to all of the epochs.
This is a small number, which proved to be insufficient to precisely
determine the value of $b$ in the fitting procedure --- the value was
large and had the wrong sign.  However, $b$ depends on the number of
charge traps as a function of time rather than on the observed field.
Thus, $b$ is derived from images of a field in Fornax, which contain
almost 200 objects and were taken within a few days of those for
UMI~J$1508+6717$.  The result is $b=-3.5\times 10^{-5}$, with the
correction applying to objects with a $S/N < 100$.
Equation~\ref{eq:cti-pc} for UMI~J$1508+6717$ has $b$ held constant at
this value and applied to objects with $S/N < 100$.  The choice for the
individual $\chi^2$ that triggers fitting for uniform motion is 30.
This results in fitting a motion for only the QSO and one star with a
large motion.  We chose not to fit a motion for any other stars because
of the small sample size, though the final fitted transformation and
motion of the QSO are not sensitive to fitting motions for a few
additional objects.

The $\chi^2$ per degree of freedom from equation~\ref{eq:k2} is again
larger than one when using the uncertainties from equation~\ref{eq:uncnew}.
Figure~\ref{fig:k2pc} plots the contribution to $\chi^2$ \textit{versus}
$S/N$ for objects in UMI~J$1508+6717$.  As in Figure~\ref{fig:k2stis},
the lack of a trend implies that the additional error is multiplicative.
A multiplier of 1.413 yields a $\chi^2$ per degree of freedom of one. 

	Figure~\ref{fig:pc-rxry} plots the position residuals $RX$ and
$RY$ for the UMI~J$1508+6717$ field, similar to Figure~\ref{fig:stis-rxry}.
No panel shows a trend of $RX$ with $X$ or $RY$ with $Y$.
Although not depicted in the figure, there are no trends between $RX$
and $Y$ or $RY$ and $X$ either.

	Figure~\ref{xqyqpc}\ shows the location of the QSO as a
function of time in the standard coordinate system for the
UMI~J$1508-6717$ field.  Note that the slopes of the corresponding
plots in Figures~\ref{xqyqpc}\ and \ref{xqyqstis}\ need not be the same
because the two fields are rotated with respect to each other.  The
motion of the QSO is $(\mu_{x},\mu_{y})=(-0.0016\pm 0.0026,-0.0085\pm
0.0028)$~pixel~yr$^{-1}$.  The contribution to the total $\chi^2$ from
the QSO is 0.01, which is reflected in the close agreement between the
points and the straight lines in Figure~\ref{xqyqpc}.  The contribution
to the $\chi^2$ has approximately two degrees of freedom, which implies
a 0.5\% probability of a $\chi^2$ smaller than 0.01 by chance.

The small contributions to the $\chi^2$ by the QSO for both the STIS
and WFPC2 data suggest that our estimated uncertainties in the
centroids are too small.  However, other objects in our fields with
fitted motions do not have unusually small $\chi^2$ values --- see
Tables~3A and 3B presented in Section~\ref{measuredpm}.  A smaller
threshold for the contribution to $\chi^2$ which triggers fitting for a
motion would reduce the multiplier needed to make the total $\chi^2$
per degree of freedom equal to one, which would increase the
contribution to the $\chi^2$ from the QSO.  We choose not to reduce our
thresholds further because the small number of stars in both of our
fields could produce a systematic error in the measured centroids
depending on position.  The small number both forces us to use an ePSF
that is constant in both space and time within an epoch and limits our
ability to detect systematic errors.  It also limits the amount of
information available to determine the transformation to the standard
coordinate system.  We prefer to average over possible systematic
errors rather than spuriously remove them by fitting motions.  We then
think that it is best to increase the uncertainties in the measured
centroids so that the uncertainty in the fitted motion of the QSO
reflects the typical scatter of the centroids around the fitted
transformation and motions --- even if the centroids of the QSO have a
smaller scatter.  We again emphasize that the measured motion of the
QSO does not depend strongly on the sample of objects with fitted
motions.

\subsection{Measured Proper Motion}
\label{measuredpm}

	Table~2 records the measured proper motion for each field in
the equatorial coordinate system and their weighted mean.  The uncertainty
in the proper motion of the dSph depends on both the uncertainty in the
measured motion of the QSO in the standard coordinate system and on the
length of the time baseline.  The baseline for the UMI~J$1508+6716$ field is
about 2~yrs, whereas that for the UMI~J$1508-6717$ field is about 4~yrs.
Because of the small number of objects in the latter field, we chose
to give the two fields more nearly equal weight in the average.  We did
this by doubling the uncertainty for the proper motion derived from
the UMI~J$1508-6717$ field and this is the uncertainty listed in Table~2.

The proper motion in Table~2 is that measured by a heliocentric
observer and, thus, includes the effects of the motions of the LSR and
of the Sun with respect to the LSR.  The measured proper motion is the
best quantity to use for comparisons with the other independent
measurements.

	There are two previous measurements of the proper motion for
Ursa Minor.  Scholz~\&~Irwin (1993) report a measured proper motion of
$(\mu_\alpha,\mu_\delta) = (50\pm 80,120\pm 80)$~mas~cent$^{-1}$ (the
last row in their Table~3, but using the larger uncertainty for
$\mu_{\delta}$ implied in the text) and Schweitzer (1996; see also
Schweitzer, Cudworth, \& Majewski 1997) reports a value of $(5.6\pm
7.8,7.4\pm 9.9)$~mas~cent$^{-1}$.  The four rectangles in
Figure~\ref{mus-comp} depict the four independent measurements of the
proper motion.  The center of a rectangle is the best estimate of the
proper motion and the sides are offset from the center by the
1-$\sigma$ uncertainties.  Rectangles 1 and 2 represent the
measurements by Scholz~\&~Irwin (1993) and Schweitzer (1996),
respectively, whereas rectangles 3 and 4 represent the measurements
from this study for the fields UMI~J$1508+6716$ and UMI~J$1508+6717$,
respectively.

Our measurements, 3 and 4, agree within their uncertainties.  The
$\mu_{\alpha}$ components of measurements 2 and 3 disagree and at least
one of these two measurements must be affected by systematic errors
larger than the quoted uncertainties.  Our other measurement, 4, is
closer to 3 than to 2.  Because of its large uncertainty, measurement 1
does not provide much additional information on the proper motion.  The
weighted mean of our measurements 3 and 4, listed in the bottom line of
Table~2, differs from measurement 2 by 3.0$\times$ the uncertainty in
the difference of $\mu_{\alpha}$ and 0.8$\times$ the uncertainty in the
difference of $\mu_{\delta}$.  Thus, these two reported measurements do
not agree within their uncertainties and one or the other or both must
have systematic errors larger than the quoted uncertainties.  Through
experimentation, we have eliminated the following sources for the
disagreement:  1) The exact shape of the ePSF.  Using an analytic ePSF
and one derived from archival data sets for a field in the globular
cluster $\omega$~Centauri had little effect on the final proper
motion.  2) The value for parameter $b$ in Equation~\ref{eq:cti-pc},
which approximately corrects for the effect of the increasing number of
charge traps in the PC2 CCD.  Large and arbitrary changes in the value
cannot simultaneously reconcile measurements 2, 3, and 4.  3) Changes
in the number of stars which determine the transformation between
epochs.  The fitted parameters, including the motion of the QSO, do not
change significantly if objects with apparently discrepant measurements
--- \textit{i.e.}, having a large $\chi^2$ after being fit for a
uniform motion --- are excluded from the sample.  4)  The exact value
of the limit on $\chi^2$ which triggers fitting for a uniform motion.
Thus, we are unable to explain the origin of the difference between our
measured proper motion and that of Schweitzer (1996; see also
Schweitzer, Cudworth, \& Majewski 1997).

Table~3A tabulates the proper motions for those objects in the
UMI~J$1508+6716$ field for which it was measured.  Table~3B does the
same for the UMI~J$1508+6717$ field.  The first line of each table
corresponds to the QSO and subsequent objects are listed in order of
decreasing $S/N$.  The ID number of an object is in column~1, the $X$
and $Y$ coordinates of an object in the earliest image of the first
epoch (o5bl01010 for UMI~J$1508+6716$ and u50j0101r for
UMI~J$1508+6717$) are in columns 2 and 3, and the $S/N$ of the object
at the first epoch is in column 4.  The components of the measured
proper motion, expressed in the equatorial coordinate system, are in
columns 5 and 6.  Each value is the measured proper motion in the
standard coordinate system corrected by adding the weighted mean proper
motion of Ursa Minor given in the bottom line of Table~2.  To indicate
that this correction has been made, the proper motion of the QSO is
given as zero.  The listed uncertainty of each proper motion is the
uncertainty of the measured proper motion, calculated in the same way
as for the QSO, added in quadrature to that of the average proper
motion of the dSph.  The contribution of the object to the total
$\chi^2$ is in column~7.  The proper motion is unreliable if this value
is much larger than 2.0.

\subsection{Galactic Rest Frame Proper Motion}
\label{grfpm}

	Removing the contributions to the measured proper motion from
the motion of the LSR and the peculiar motion of the Sun, yields the
Galactic rest frame proper motion.  This proper motion would be
measured by a hypothetical observer at the location of the Sun and at
rest with respect to the Galactic center.  Columns (2) and (3) in
Table~4 list the components of the Galactic rest frame proper motion
expressed in the equatorial coordinate system,
$(\mu_{\alpha}^{\mbox{\tiny{Grf}}},\mu_{\delta}^{\mbox{\tiny{Grf}}})$,
for our two fields.  The derivation of
$(\mu_{\alpha}^{\mbox{\tiny{Grf}}},\mu_{\delta}^{\mbox{\tiny{Grf}}})$
assumes: 220~km~s$^{-1}$ for the circular velocity of the LSR; 8.5~kpc
for the distance of the Sun from the Galactic center; and $(u_\odot,
v_\odot, w_\odot) = (-10.00\pm 0.36, 5.25\pm 0.62,
7.17\pm 0.38)$~km~s$^{-1}$ (Dehnen \& Binney 1998) for the peculiar
velocity of the Sun, where the components are positive if $u_{\odot}$
points radially away from the Galactic center, $v_{\odot}$ is in the
direction of rotation of the Galactic disk, and $w_\odot$ points in the
direction of the North Galactic Pole.  For convenience, columns (4) and
(5) in Table~4 list the components of the Galactic rest frame proper
motion in the Galactic coordinate system,
$(\mu_{l}^{\mbox{\tiny{Grf}}},\mu_{b}^{\mbox{\tiny{Grf}}})$.  The next
three columns list the components of the space velocity in the
cylindrical coordinate system centered on the dSph.  The derivation of
these assumes a heliocentric distance of 76~kpc to Ursa Minor.  The
components of the space velocity are positive if $\Pi$ points radially
away from the Galactic rotation axis; $\Theta$ is in the direction of
rotation of the Galactic disk; and $Z$ points in the direction of the
North Galactic Pole.  The final two columns list the radial and
tangential components of the space velocity with respect to a
hypothetical observer at rest at the Galactic center.  Column (9) gives
the radial component, which is positive if it points away from the
Galactic center and column (10) gives the tangential component.

\section{Orbit and Orbital Elements of Ursa Minor}
\label{orbitsec}

	Knowing the space velocity of a dSph permits a determination
of its orbit for a given form of the Galactic potential.  This study
adopts a Galactic potential that has a contribution from a disk of the
form (Miyamoto \& Nagai 1975)
\begin{equation}
\label{diskpot}
\Psi_{\mbox{\small{disk}}}=-\frac{G
M_{\mbox{\small{disk}}}}{\sqrt{R^{2}+(a+\sqrt{Z^{2}+b^{2}})^{2}}},
\end{equation}
from a spheroid of the form (Hernquist 1990)
\begin{equation}
\label{spherpot}
\Psi_{\mbox{\small{spher}}}=-\frac{GM_{\mbox{\small{spher}}}}
{R_{\mbox{\small{GC}}}+c},
\end{equation}
and from a halo of the form
\begin{equation}
\label{logpot}
\Psi_{\mbox{\small{halo}}}=v^{2}_{\mbox{\small{halo}}}\ln
(R^{2}_{\mbox{\small{GC}}}+d^{2}).
\end{equation}
In the above equations, $R_{\mbox{\small GC}}$ is the Galactocentric
distance, $R$ is the projection of $R_{\mbox{\small GC}}$ onto the
plane of the Galactic disk, and $Z$ is the distance from the plane of
the disk.  All other quantities in the equations are adjustable
parameters and their values are the same as those adopted by Johnston,
Sigurdsson, \& Hernquist (1999):
$M_{\mbox{disk}}=1.0\times10^{11}$~M$_{\odot}$,
$M_{\mbox{spher}}=3.4\times10^{10}$~M$_{\odot}$,
$v_{\mbox{halo}}=128$~km~s$^{-1}$, $a=6.5$~kpc, $b=0.26$~kpc,
$c=0.7$~kpc, and $d=12.0$~kpc.

	Figure~\ref{orbit} shows the projections of the orbit of Ursa
Minor resulting from an integration of the motion in the Galactic
potential given by Equations~\ref{diskpot}, \ref{spherpot}, and
\ref{logpot}.  The integration extends for 3~Gyr backwards in time and
begins at the current location of Ursa Minor with the negative of the
space velocity given in the bottom line of Table~4.  The solid square
marks the current location of the dSph, the solid star indicates the
center of the Galaxy, and the small open circles mark the points where
$Z=0$ or, in other words, where the orbit crosses the plane of the
Galactic disk.  The large open circle is for reference: it has a radius
of 30~kpc.  In the right-handed coordinate system of
Figure~\ref{orbit}, the current location of the Sun is on the positive
$X$-axis.  The figure shows that Ursa Minor is close to apogalacticon
and has both a moderately inclined and eccentric orbit.

	Table~5 enumerates the elements of the orbit of Ursa Minor.
The value of the quantity is in column (4) and its $95\%$ confidence
interval is in column (5).  The latter comes from 1000 Monte Carlo
experiments, where an experiment integrates the orbit using an initial
velocity that is chosen randomly from a Gaussian distribution whose
mean and standard deviation are the best estimate of the space velocity
and its quoted uncertainty, respectively.
The eccentricity of the orbit, defined as
\begin{equation}
\label{eccentricity}
e = \frac{(R_{a} - R_{p})}{(R_{a} + R_{p})},
\end{equation}
is 0.39, though the 95\% confidence interval ranges from a nearly circular
orbit to a nearly radial orbit with a perigalacticon near 10~kpc .  The inclination, $\Phi$, implies a retrograde orbit.  The inclination of the
nominal orbit is about 56~degrees to the Galactic plane, though a nearly
polar orbit is within the 95\% confidence interval.  The longitude of the
ascending node, $\Omega$, is measured counter-clockwise from the positive
$X$-axis.

\section{Discussion}
\label{sec:disc}

\subsection{Is Ursa Minor a Member of a Stream?}
\label{stream}

	As outlined in Section~\ref{intro}, Ursa Minor may be a member
of a stream that also includes the LMC, SMC, Draco, and, possibly, Sculptor
and Carina.  If Ursa Minor is a member, then its predicted measured
proper motion is $(\mu_{\alpha},\mu_{\delta}) =
(-8,13)$~mas~cent$^{-1}$, or $\vert \vec{\mu} \vert = \sqrt
{\mu_{\alpha}^{2} + \mu_{\delta}^{2}} = 15$~mas~cent$^{-1}$ with a
position angle of 328~degrees (Lynden-Bell \& Lynden-Bell 1995).
The measured proper motion from this
study is $\vert \vec{\mu} \vert = 55 \pm 17$~mas~cent$^{-1}$ with a
position angle of $294 \pm 17$~degrees, which are $2.4~\sigma$ and
$2.0~\sigma$ away from the predicted values, respectively.  We rule out
the possibility that Ursa Minor is a member of the proposed stream
at more than $2~\sigma$.

Kroupa, Theis, \& Boily (2004) show that the 11 dwarf galaxies nearest
to the Milky Way are nearly on a plane, whose pole is at $(\ell,b) =
(168,-16)$~degrees.  Adopting the direction of the angular momentum
vector as the pole of the orbit, then the location of the pole is
\begin{equation}
(\ell,b) = (\Omega+90^{\circ},\Phi-90^{\circ}).
\end{equation}
Because of the left-handed nature of the Galactic rotation, prograde
orbits have $b < 0$ and retrograde orbits have $b > 0$.  Thus, the
pole of our orbit for Ursa Minor is $(\ell,b) = (243\pm 20, 34\pm
11)$~degrees, where the uncertainties are 1-$\sigma$ values from the
Monte Carlo simulations.  Thus, the motion of Ursa Minor is not in the
plane defined by the nearby dwarf galaxies.

\subsection{Star Formation History in Ursa Minor}

	Studies of the stellar population in Ursa Minor, described in
Section~\ref{intro}, indicate that the dSph contains low-metallicity,
old stars.  The majority of stars formed close to a Hubble time ago,
suggesting that Ursa Minor lost its gas quickly.  In contrast, Carina
had extensive star formation about 7~Gyr ago, which continued to within
1~Gyr ago.  These two galaxies have nearly the same luminosity and
surface brightness and, thus, presumably should have retained gas to a
similar degree.  Since they have not, is it because their Galactic
orbits are different?  The answer appears to be no.

Table~4 of P03 shows that Carina has orbital elements similar to those
of Ursa Minor.  The perigalacticon of Carina is probably smaller than
that of Ursa Minor (a nominal value of 20~kpc \textit{versus}\ 40~kpc),
so Carina should have lost its gas more rapidly to stronger tidal
shocks (Mayer \etal\ 2001) and larger ram pressure stripping (Blitz \&
Robishaw 2002; Gallart \etal\ 2001; Mayer \& Wadsley 2003).  A notable
difference between the two orbits is that that for Carina is prograde,
while that for Ursa Minor is retrograde.  However, how this difference
could have affected star formation is unclear.

\subsection{A Lower Limit for the Mass of the Milky Way}
\label{massofg}

	Ursa Minor is bound gravitationally to the Milky Way.  The
Galactocentric space velocity of the dSph imposes a lower limit on the
mass of the Milky Way within the present Galactocentric radius of the
dSph, $R$.  Assuming a spherically symmetric mass distribution and zero
for the total energy of the dSph, the lower limit for the mass
of the Milky Way is given by
\begin{equation}
\label{mwmass}
M=\frac{R\left (V_{r}^{2} + V_{t}^{2} \right )}{2G}.
\end{equation}
Setting $R=78$~kpc and using the values from Table~4 for $V_{r}$ and
$V_{t}$, $M = (2.4 \pm 1.4) \times 10^{11}\ M_{\odot}$.  This lower limit
is consistent with other recent estimates of the mass of the Milky
Way, such as the mass of $5.4^{+0.1}_{-0.4} \times 10^{11}\ M_{\odot}$
within $R=50$~kpc found by Sakamoto, Chiba, \& Beers (2003).  The
Milky Way potential adopted in Section~\ref{orbitsec} has a mass of
$7\times 10^{11}\ M_{\odot}$ out to $R=78$~kpc.

\subsection{The Effect of the Galactic Tidal Force on Structure of
Ursa Minor}

	The measured ellipticity of Ursa Minor is one of the
largest known for the dSphs.  If the Galactic tidal force deformed
Ursa Minor from an initial spherical shape to its present elongated
shape, then the position angle of its projected major axis should be
similar to the position angle of the Galactic-rest-frame proper motion
vector.  The position angle of the projected major axis is $53 \pm
5$~degrees and the position angle of the Galactic-rest-frame proper
motion vector is $348 \pm 25$~degrees.  The difference between the two
position angles is 2.6 times its uncertainty, arguing that Ursa Minor
is not elongated along its orbit.

G\'{o}mez-Flechoso \& Mart\'{\i}nez-Delgado (2003) derive an $M/L_V$
for Ursa Minor in the range 6 to 24 by matching the radial profile of a
model dSph from N-body simulations to the observed radial profile.  In
the simulations, the dSph moves on the orbit found by Schweitzer,
Cudworth, \& Majewski (1997).  Given our different orbit for Ursa
Minor, is this range of $M/L_V$ values still compatible with the
observed limiting radius of the radial profile?  A poor-man's
substitute for numerical simulations is to calculate the tidal radius,
$r_t$, beyond which a star becomes unbound from the dSph.  For a
logarithmic Galactic potential, $r_t$ is given by (King 1962; Oh, Lin,
\& Aarseth 1992)
\begin {equation}
r_t = \left(\frac{(1-e)^2}{[(1+e)^2/2e]\ln[(1+e)/(1-e)] +1} \,
\frac{M}{M_g}\right)^{1/3} a.
\label{eq:rtidal}
\end {equation}
Here $e$ is the eccentricity of the orbit, $a$ is the semi-major axis ($a
\equiv (R_{a}+R_{p})/2$), $M$ is the mass of the dSph, and $M_g$ is the
mass of the Galaxy within $a$.  Equating $r_t$ with the observed
limiting radius derived by fitting a King (1966) model, $r_k$, yields a
value for $M/L_V$ for a given orbit.  If $r_k = 50$~arcmin, then 50\%
of the orbits in Monte Carlo simulations have $M/L_V > 24$.  This
$M/L_V$ is just within the range quoted by G\'{o}mez-Flechoso \&
Mart\'{\i}nez-Delgado (2003).  If $r_k = 78$~arcmin, then 50\% of the
orbits have $M/L_V > 89$, and 95\% have $M/L_V > 24$.  Thus, the
$M/L_V$ derived by G\'{o}mez-Flechoso \& Mart\'{\i}nez-Delgado (2003)
is incompatible with the larger measured value of $r_k = 78$~arcmin for
our measured proper motion for Ursa Minor.  On the other hand, any of
the larger values for $M/L_V$ derived from the observed velocity
dispersion (see Section~\ref{intro}) is more compatible with the larger
value of $r_k$.  Equation~\ref{eq:rtidal}\ shows that $M \propto
r_t^3$, so the values for $M/L$ derived using this equation are
sensitive to the measured value of the limiting radius and the
identification of that radius with the tidal radius.  Until kinematic
measurements definitively identify the tidal radius, an $M/L$ derived
with the above argument should be treated with caution.

The average measured $M/L_V$ for Galactic globular clusters is 2.3
(Pryor \& Meylan 1993).  Could the $M/L_V$ of Ursa Minor be this low?
Numerical simulations by Oh, Lin, \& Aarseth (1995) and Piatek \& Pryor
(1995) show that the ratio of the limiting radius derived by fitting a
theoretical King model (King 1966), $r_k$, to the tidal radius defined
by Equation~(\ref{eq:rtidal}) is a useful indicator of the importance
of the Galactic tidal force on the structure of a dSph.  These
simulations show that: if $r_{k}/r_t \lesssim 1.0$, the Galactic tidal
force has little effect on the structure of the dSph; at $r_k/r_t
\approx 2.0$, the effect of the force increases rapidly with increasing
$r_k/r_t$; and, for $r_k/r_t \approx 3.0$, the dSph disintegrates in a
few orbits.  Assuming that $M/L_V = 2.3$ and $r_k = 50$~arcmin,
$r_k/r_t > 2.0$ for 59\% of the orbits generated in Monte Carlo
simulations.  If $r_k = 78$~arcmin, the fraction is 99.6\%.  Thus, it
is unlikely that Ursa Minor would have survived a Hubble time on its
current orbit if it did not contain any dark matter.

\section{Summary}
\label{sec:summary}

1) Two independent measurements of the proper motion for Ursa Minor
produce a weighted-average value of $(\mu_{\alpha},\mu_{\delta})=(-50
\pm 17,22 \pm 16)$~mas~century$^{-1}$ in the equatorial coordinate system
for a heliocentric observer.  Our value and the proper motion of 
Schweitzer (1996; also Schweitzer, Cudworth, \& Majewski 1997) disagree
by more than twice the uncertainty in the difference.

2) Removing the contributions of the motion of the Sun and of the LSR
to the measured proper motion, gives a Galactic-Rest-Frame proper
motion of
$(\mu_{\alpha}^{\mbox{\tiny{Grf}}},\mu_{\delta}^{\mbox{\tiny{Grf}}})=(-8
\pm 17, 38 \pm 16)$ mas~century$^{-1}$ in the equatorial coordinate
system for an observer at the location of the Sun but at rest with
respect to the Galactic center.  In the Galactic coordinate system,
this motion is
$(\mu_{l}^{\mbox{\tiny{Grf}}},\mu_{b}^{\mbox{\tiny{Grf}}}) =
(32 \pm 17,-22 \pm 17)$ mas~century$^{-1}$.

3) For an observer located at the Galactic center and at rest, the
radial and tangential components of the space velocity are
$V_{r}=-75\pm 44$ km~s$^{-1}$ and $V_{t}=144 \pm 50$ km~s$^{-1}$,
respectively.

4) The best estimate for the orbit shows that Ursa Minor is close to
its apogalacticon of $R_{a}=89$ kpc and is moving closer to the Milky
Way on a retrograde orbit with eccentricity $e=0.39$ and a smallest
angle between the orbital plane and the plane of the Galactic disk of
$56^{\circ}$.  The closest approach to the Galaxy, perigalacticon, is
$R_{p}=40$ kpc.  Ursa Minor completes one full circuit around the Milky
Way in $T=1.5$~Gyr.

5) Ursa Minor is not a likely member of the ``stream'' of galaxies on
similar orbits proposed by Lynden-Bell \& Lynden-Bell (1995), nor is
its orbit confined to the plane of satellite galaxies noted by Kroupa,
Theis, \& Boily (2004).

6)  Excluding the possibility of exotic physics, \textit{e.g.}, MOND
(Milgrom 1983), Ursa Minor must contain dark matter to have a high
probability of surviving for a Hubble time on its current orbit.

\acknowledgments

We thank Daniel Eisenstein and Jim Liebert for obtaining new spectra
for the QSOs in Ursa Minor after the discovery spectra were lost.  
We thank the referee, Dr.\ van~Altena, for helpful suggestions.  CP
and SP acknowledge the financial support of the Space Telescope Science
Institute through the grants HST-GO-07341.03-A and HST-GO-08286.03-A.
EWO acknowledges support from the Space Telescope Science Institute
through the grants HST-GO-07341.01-A and HST-GO-08286.01-A and from the
National Science Foundation through the grants AST-9619524 and
AST-0098518.  MM acknowledges support from the Space Telescope Science
Institute through the grants HST-GO-07341.02-A and HST-GO-08286.02-A
and from the National Science Foundation through the grant
AST-0098661.  DM is supported by FONDAP Center for Astrophysics
15010003.

\clearpage

\setcounter{figure}{0}
\begin{figure}
\includegraphics[angle=-90,scale=0.7]{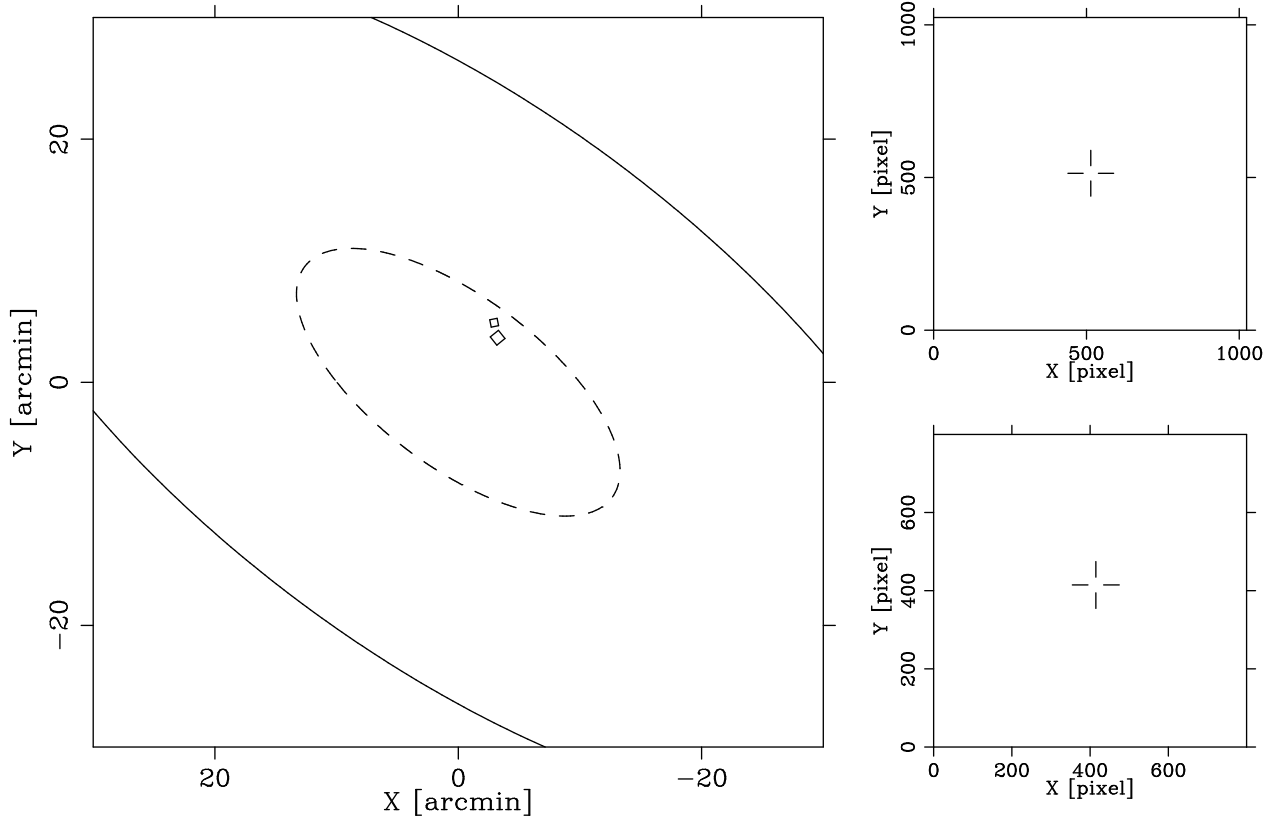}
\caption{Left panel: An image of the sky in the direction of the Ursa
Minor dSph.  The dashed ellipse is the measured core radius and the
solid ellipse is the measured tidal boundary.  The two squares
represent the fields studied in this article.  The larger of the two
corresponds to the UMI~J$1508+6716$ field and the smaller to the
UMI~J$1508+6717$ field.  Top-right panel: A sample image from the
epoch 2000 data for the UMI~J$1508+6716$ field.  The cross-hair
indicates the location of the QSO.  Bottom-right panel: A sample image
from the epoch 1999 data for the UMI~J$1508+6717$ field.  Again, the
cross-hair indicates the location of the QSO.}
\label{fields}
\end{figure}
\clearpage
\begin{figure}
\includegraphics[angle=-90,scale=1.0]{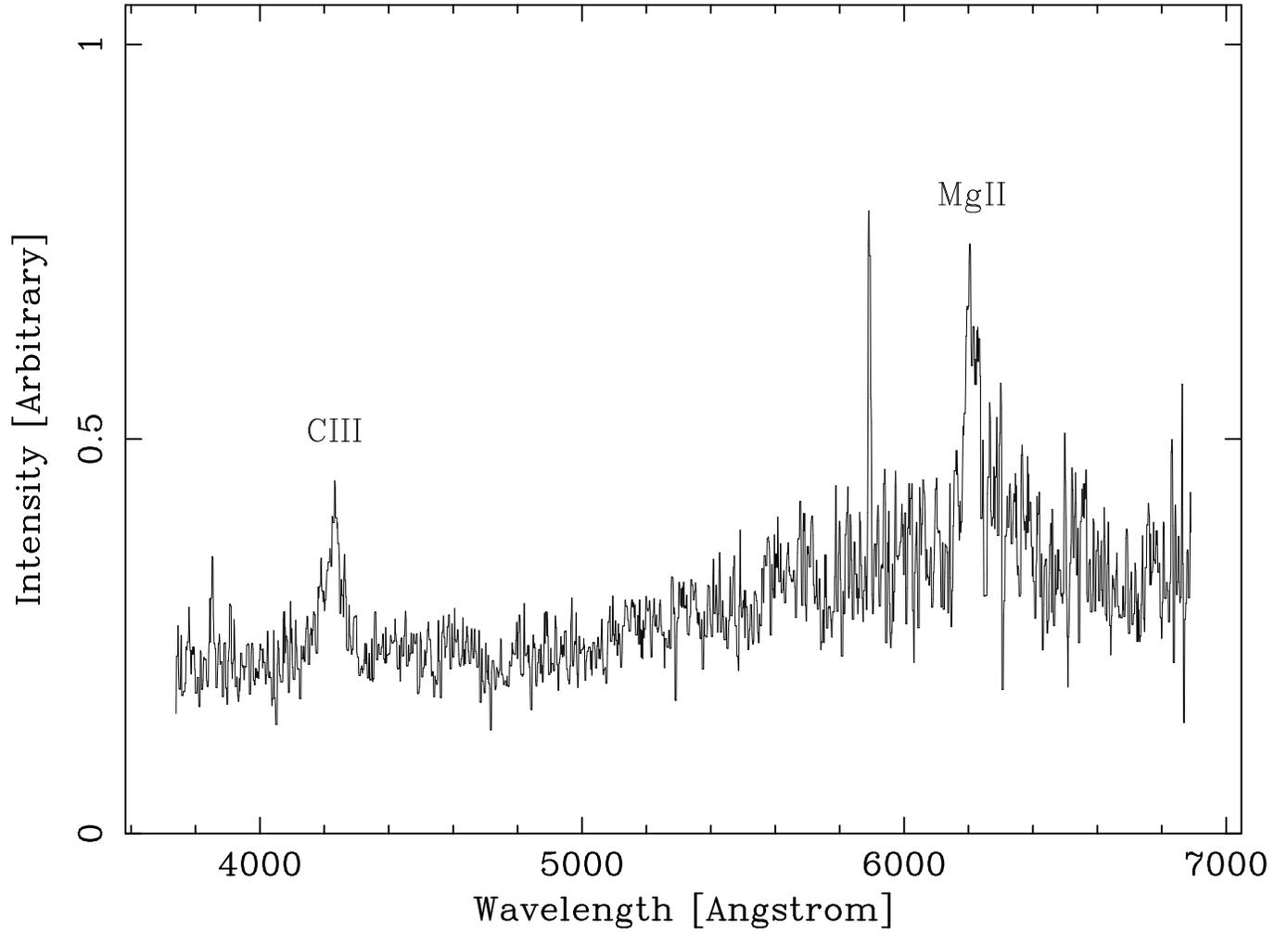}
\caption{Spectrum of the QSO in the UMI~J$1508+6716$ field taken with
the Blue Channel Spectrograph and the 500-line grating on the Multiple
Mirror Telescope on January 19, 2004.  This setup gives a resolution of
3.6~\AA\ and a coverage of 3200~\AA.  The spectrum is the sum of two
600~s exposures and it has been smoothed by a running median of five
points.  The measured redshift is 1.216.}
\label{fig:QSO-STIS}
\end{figure}
\clearpage
\begin{figure}
\includegraphics[angle=-90,scale=1.0]{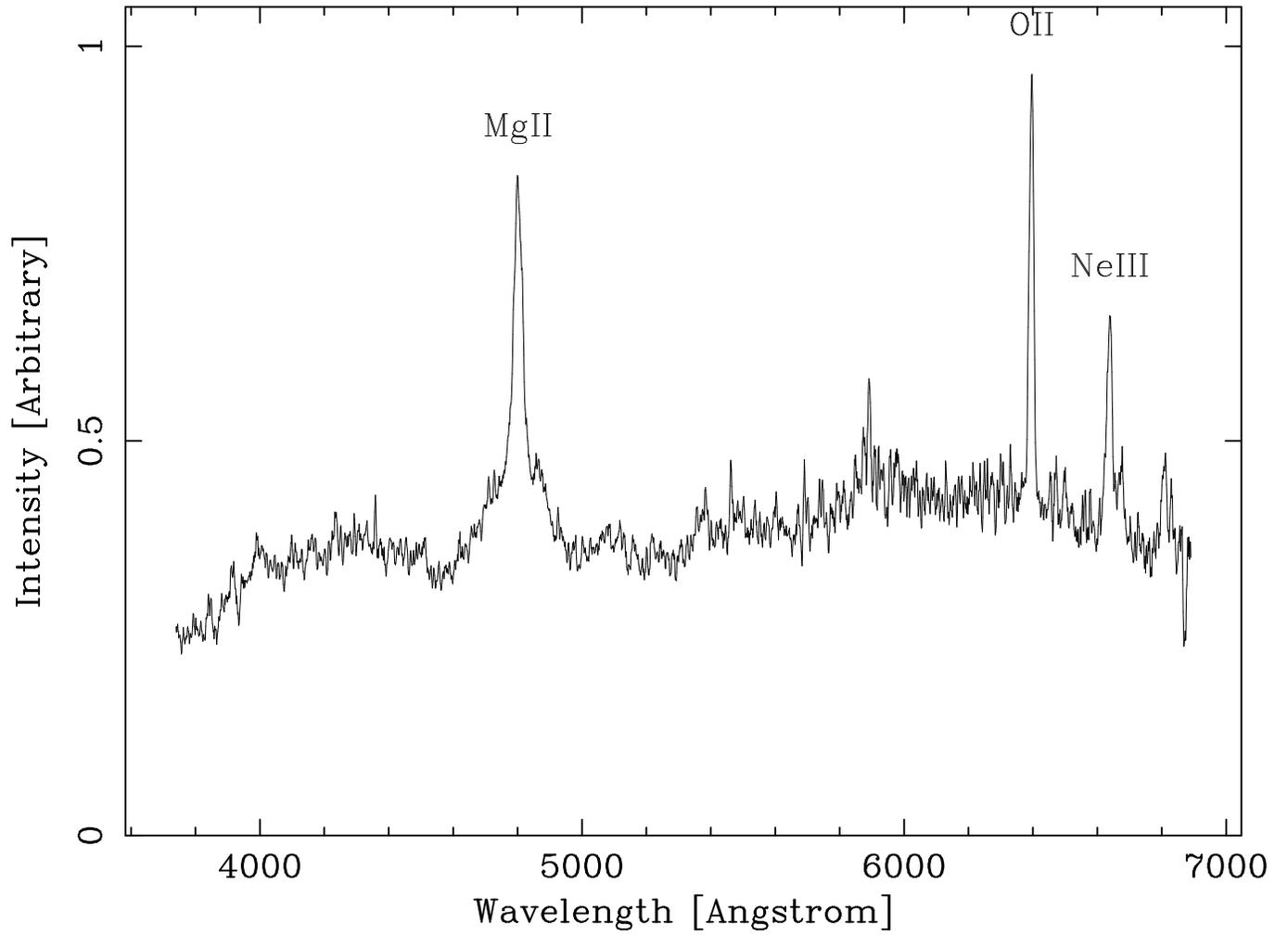}
\caption{The same as Figure~2 for the QSO in the UMI~J$1508+6717$
field.  This single 600~s exposure has been smoothed with a running
median of five points.  The measured redshift is 0.716.}
\label{fig:QSO-WFPC}
\end{figure}
\clearpage
\begin{figure}
\centering
\subfigure[]
{\label{stis-8-rf-x}
\includegraphics[angle=-90,scale=0.28]{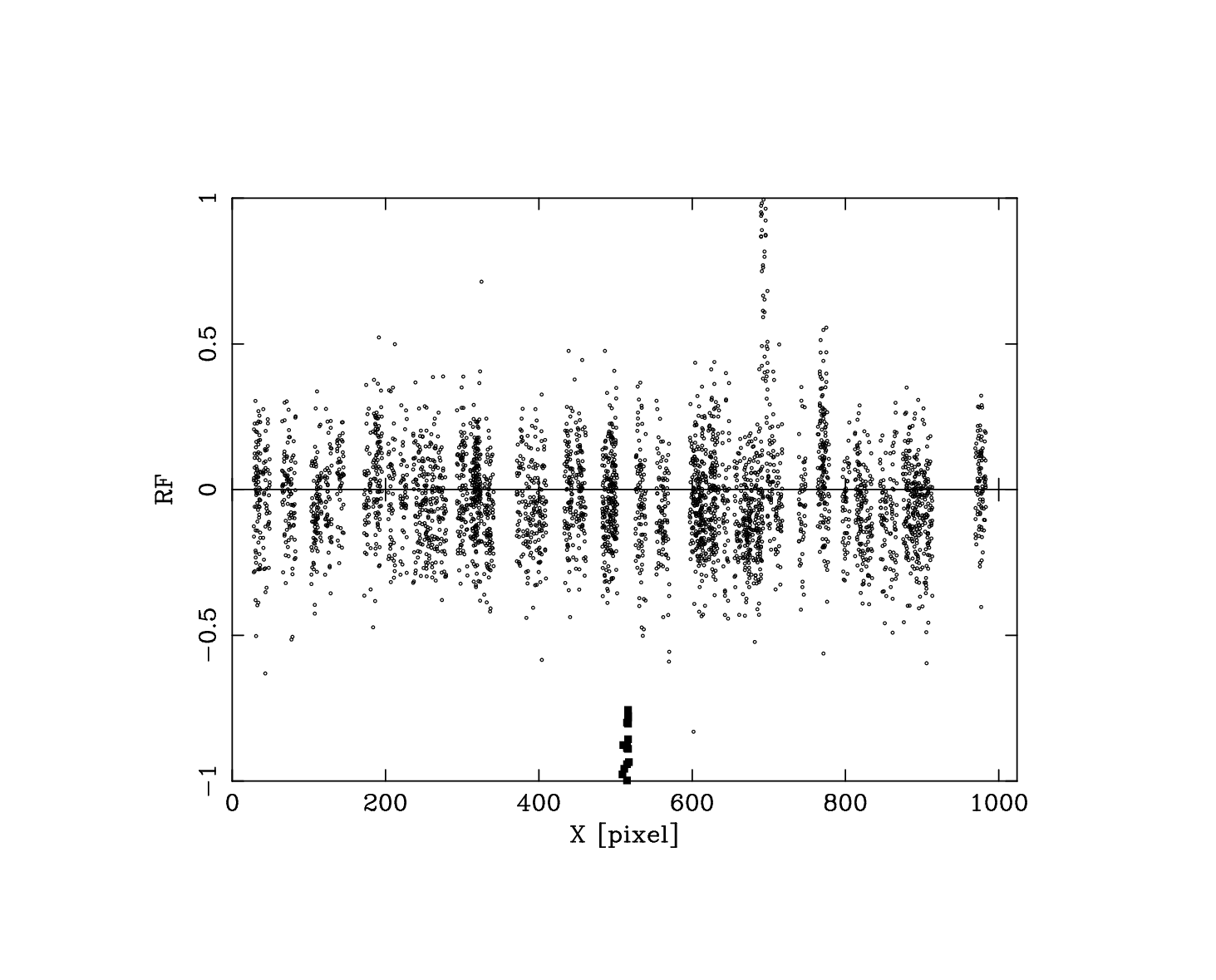}
}
\subfigure[]
{\label{stis-8-rf-y}
\includegraphics[angle=-90,scale=0.28]{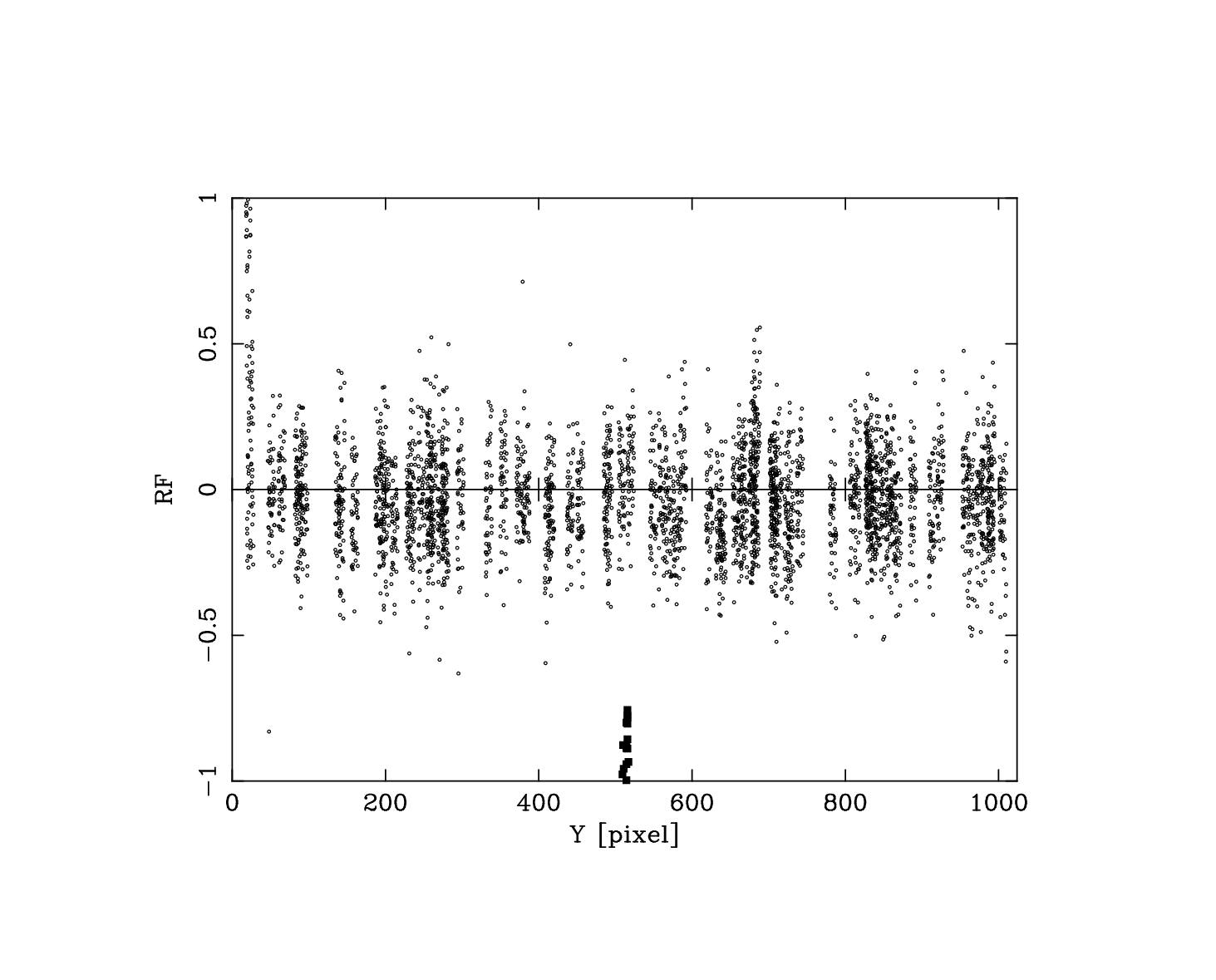}
}
\subfigure[]
{\label{stis-9-rf-x}
\includegraphics[angle=-90,scale=0.28]{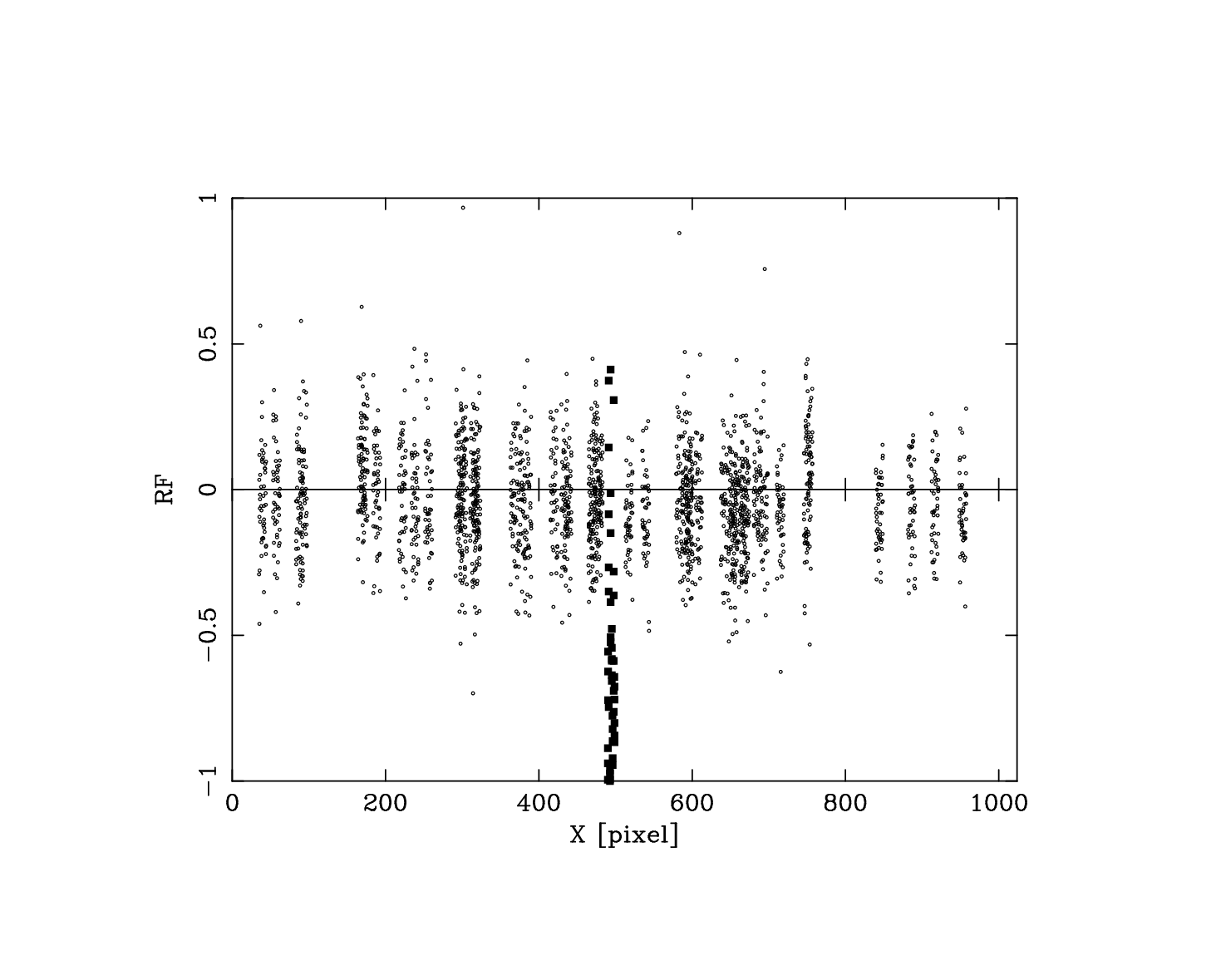}
}
\subfigure[]
{\label{stis-9-rf-y}
\includegraphics[angle=-90,scale=0.28]{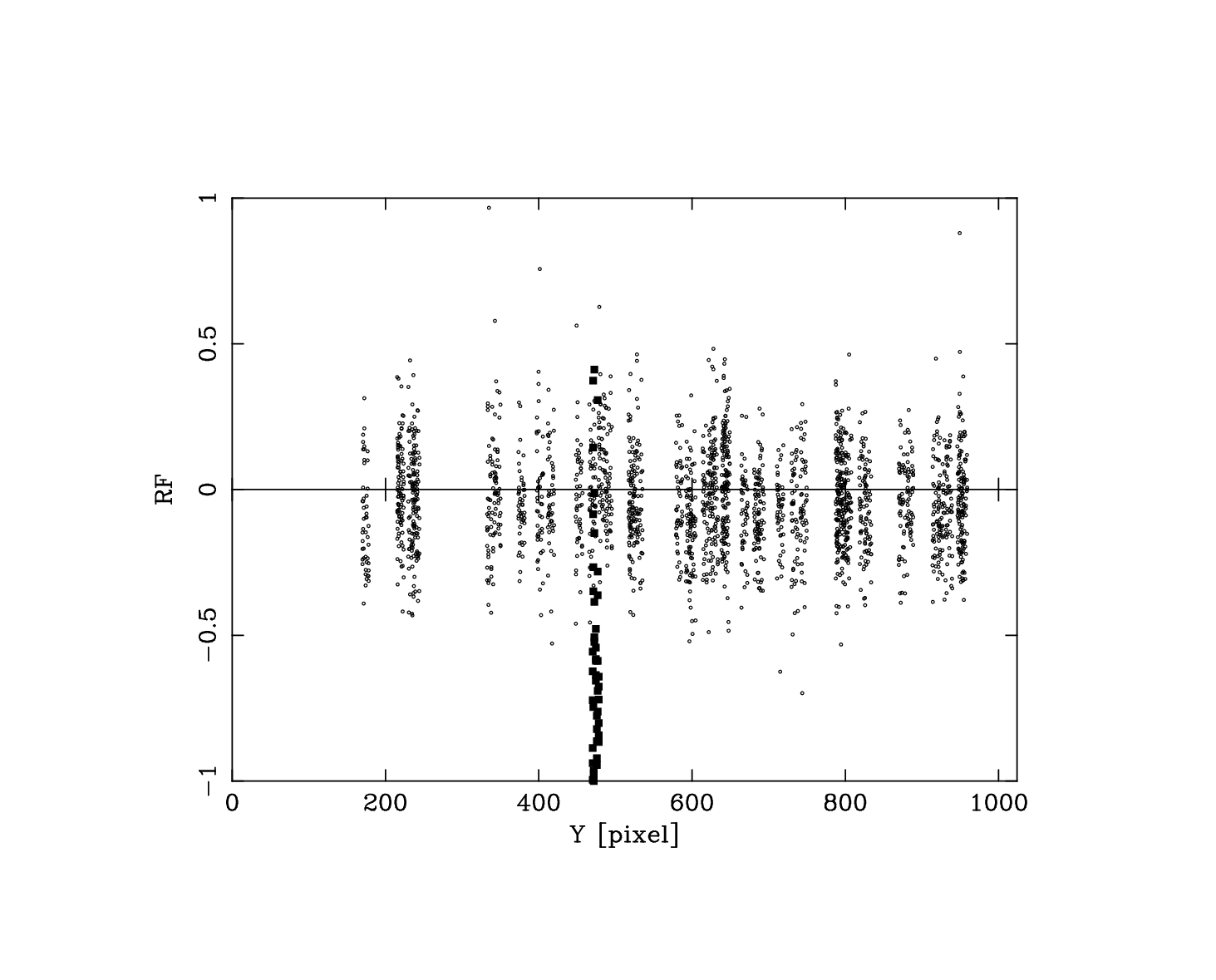}
}
\subfigure[]
{\label{stis-10-rf-x}
\includegraphics[angle=-90,scale=0.28]{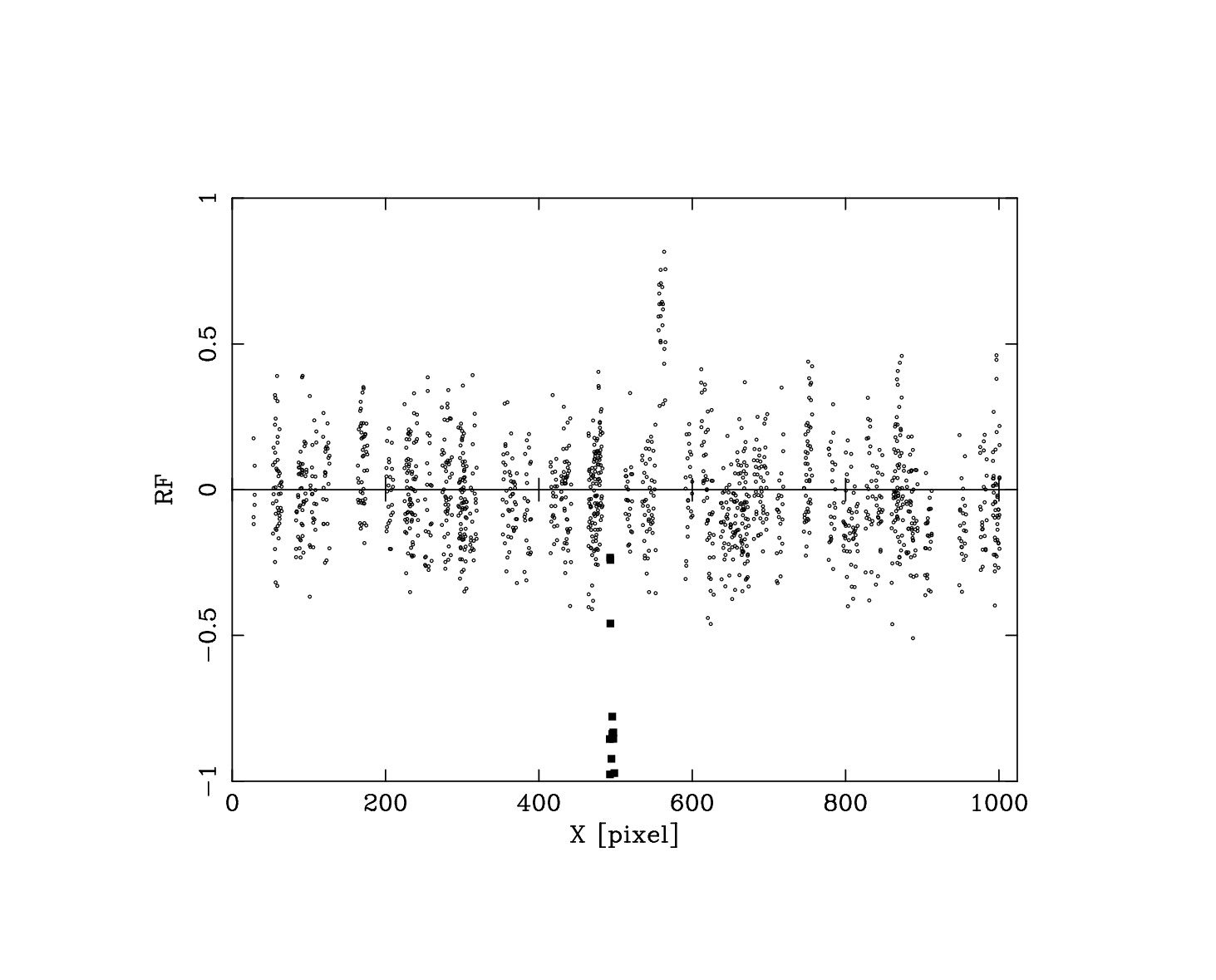}
}
\subfigure[]
{\label{stis-10-rf-y}
\includegraphics[angle=-90,scale=0.28]{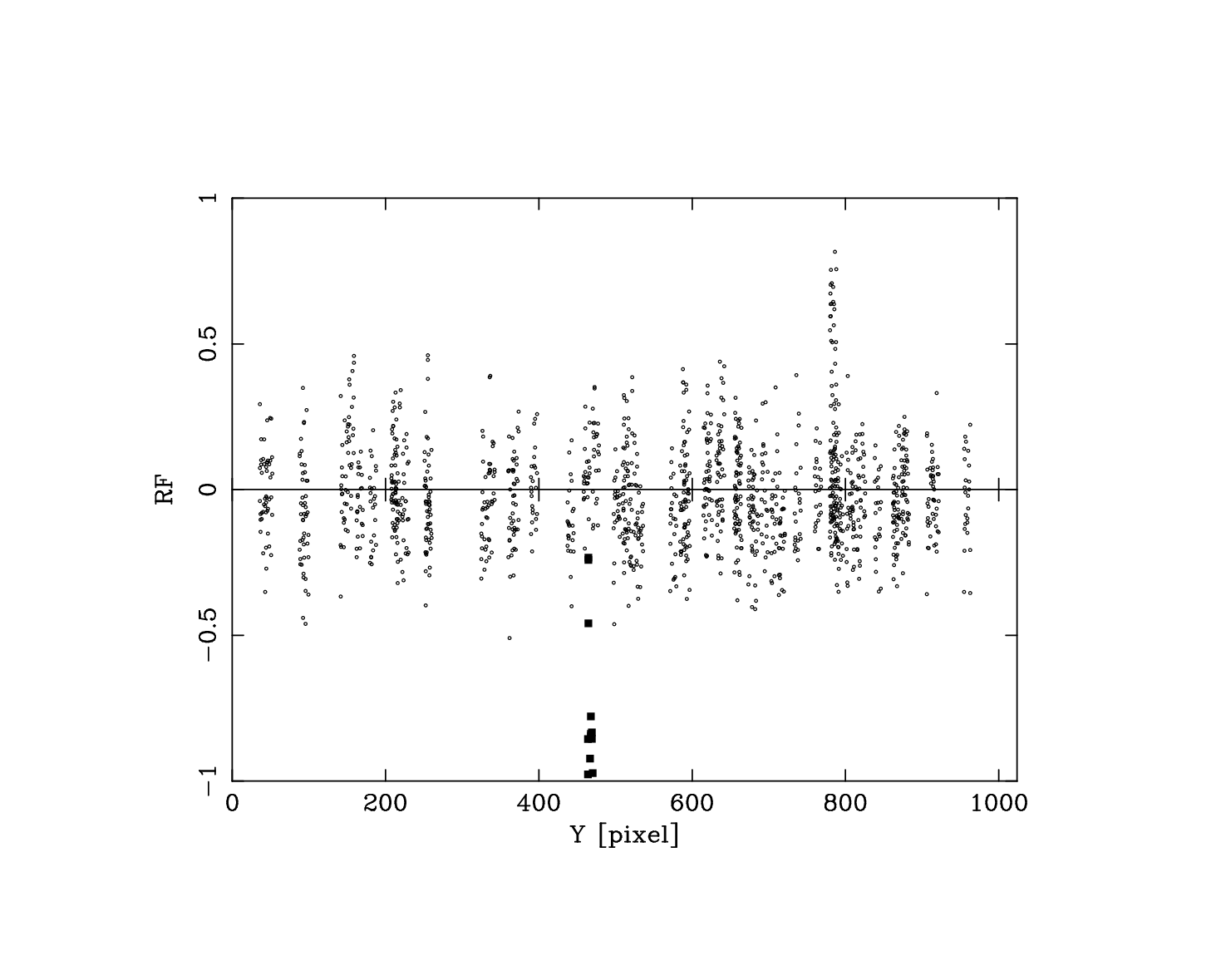}
}
\caption{Flux residual \textit{versus} location for objects in an
image for the UMI~J$1508+6716$ field.  A plot in the top two rows
displays points from 48 images and a plot in the remaining row
displays points from 24 images.  The left panels plot ${\cal RF}$
\textit{versus} the $X$-coordinate and the right panels plot ${\cal
RF}$ \textit{versus} the $Y$-coordinate.  The solid square represents
the QSO.  Panels a) and b) correspond to the 2000 epoch; c) and d) to
the 2001 epoch; and e) and f) to the 2002 epoch.  For ease of
comparison, all of the plots have the same scale on the vertical axis.}
\label{rf-stis}
\end{figure}
\clearpage
\begin{figure}
\centering
\subfigure[]
{\label{pc-7-rf-x}
\includegraphics[angle=-90,scale=0.28]{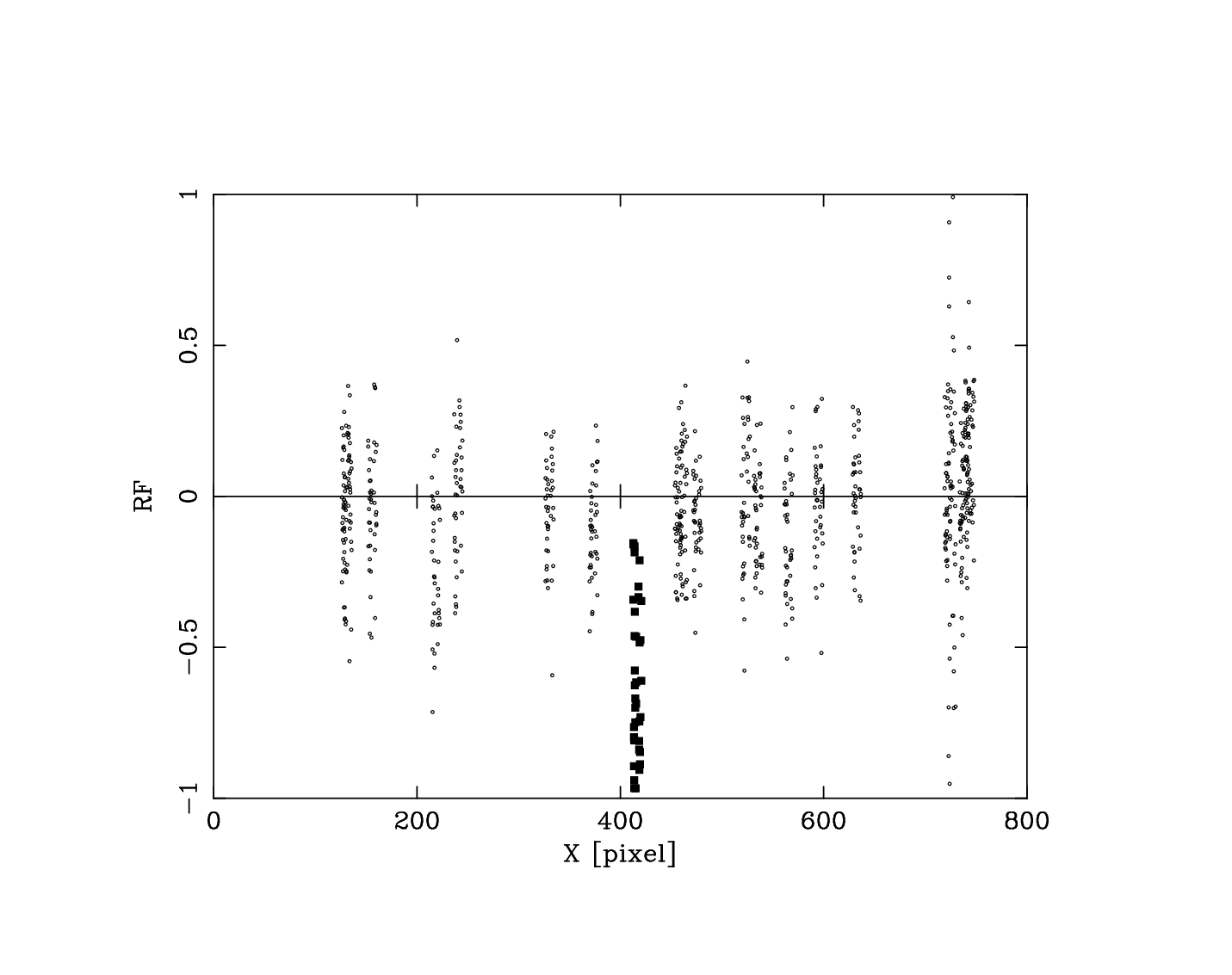}
}
\subfigure[]
{\label{pc-7-rf-y}
\includegraphics[angle=-90,scale=0.28]{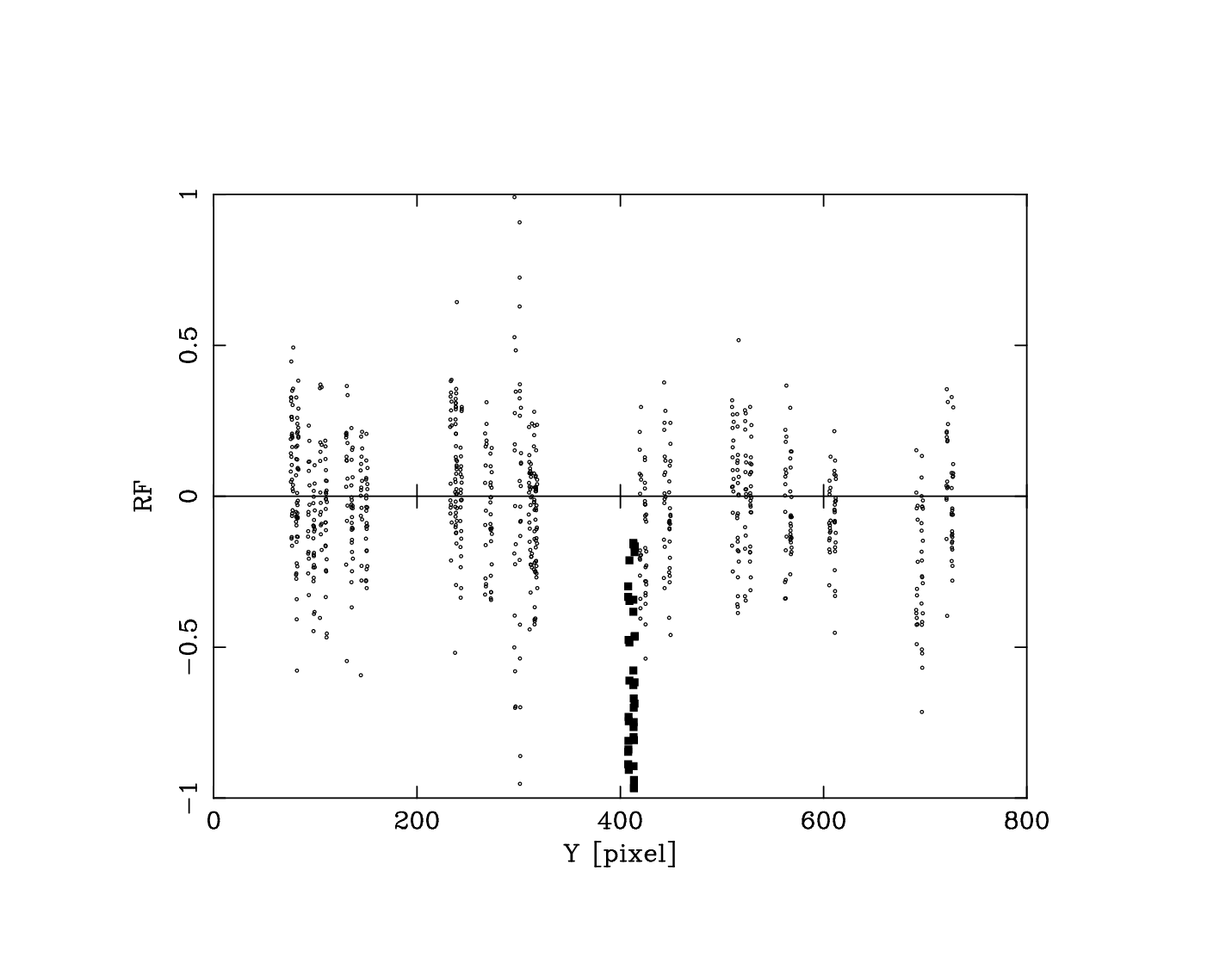}
}
\subfigure[]
{\label{pc-9-rf-x}
\includegraphics[angle=-90,scale=0.28]{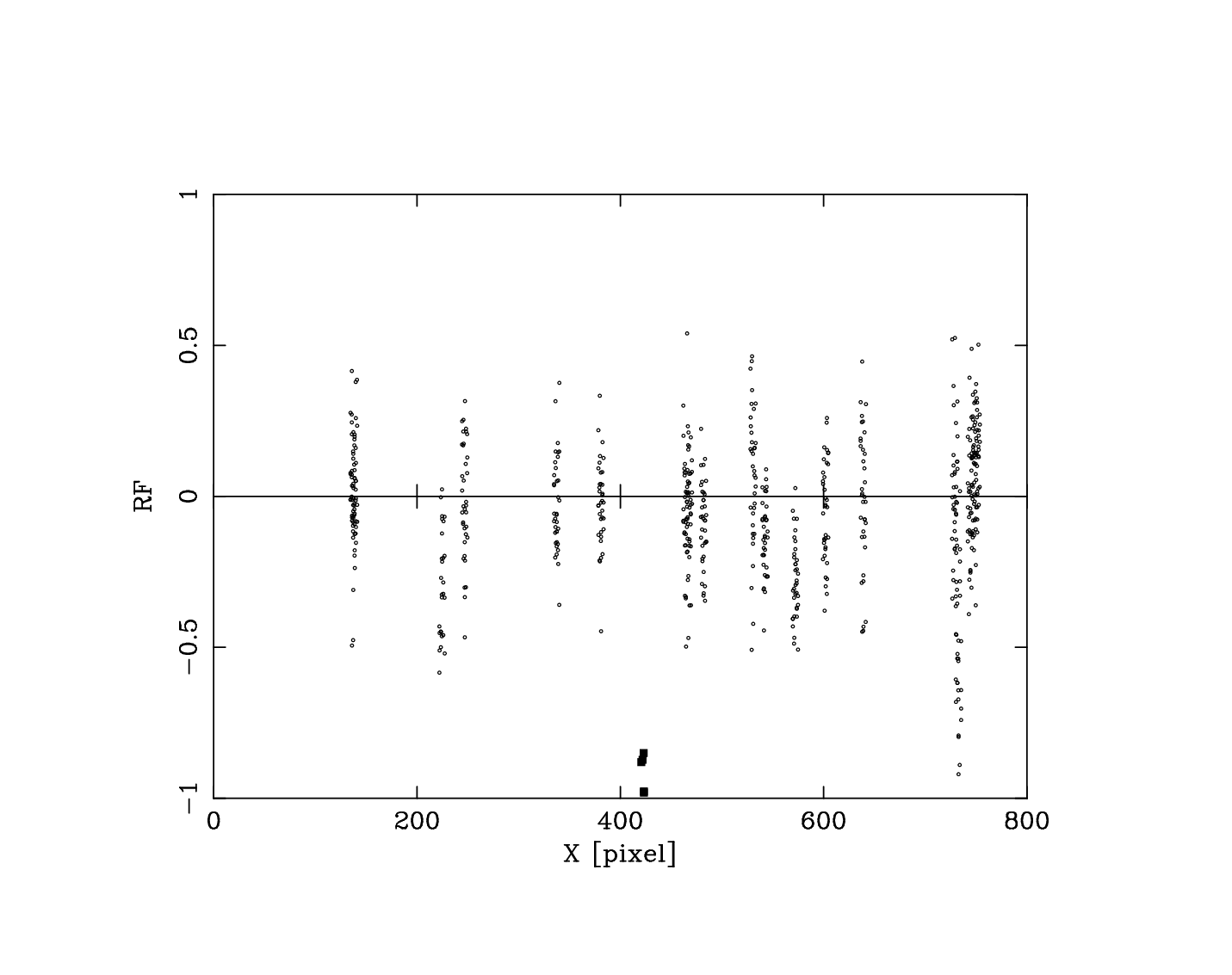}
}
\subfigure[]
{\label{pc-9-rf-y}
\includegraphics[angle=-90,scale=0.28]{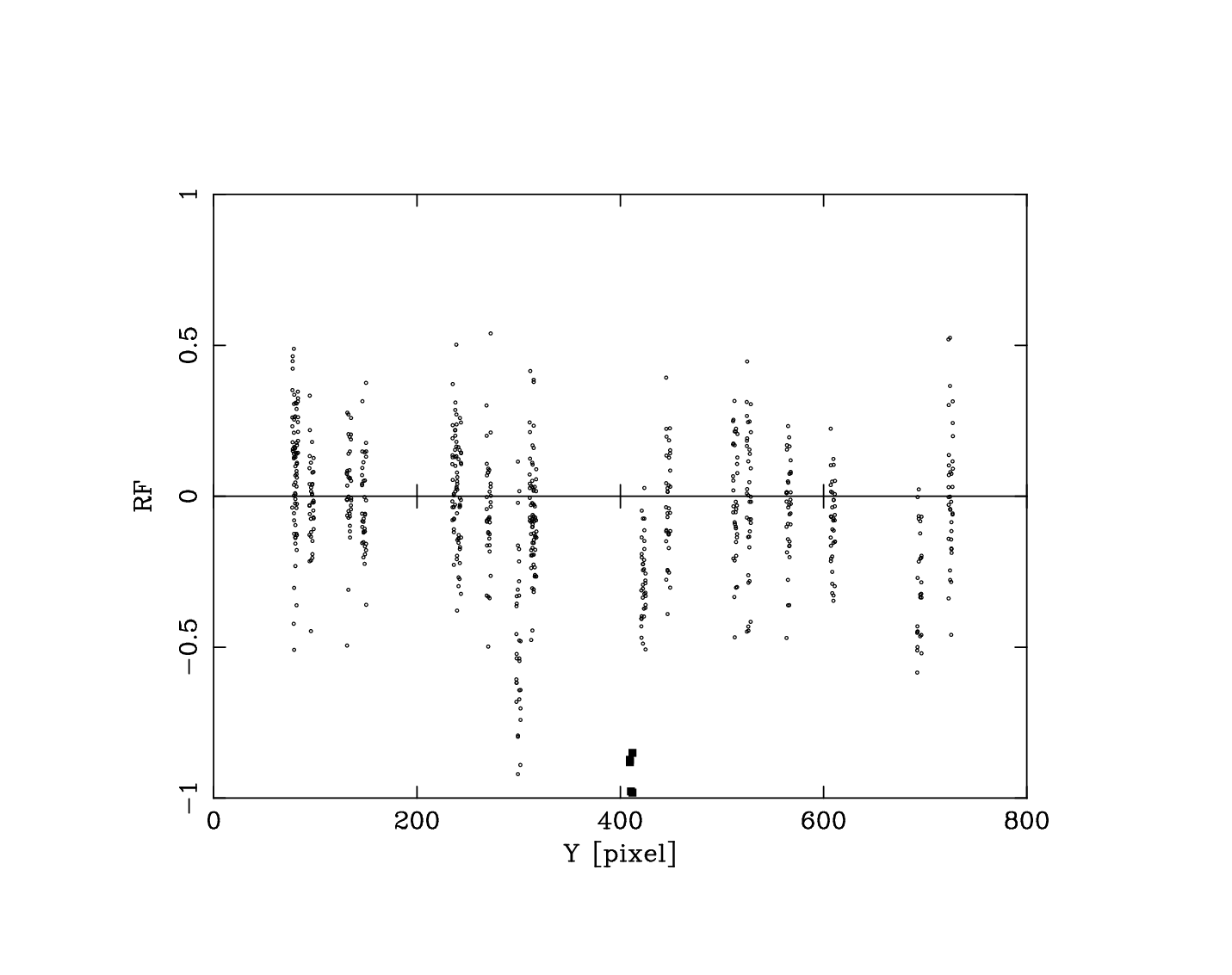}
}
\subfigure[]
{\label{pc-10-rf-x}
\includegraphics[angle=-90,scale=0.28]{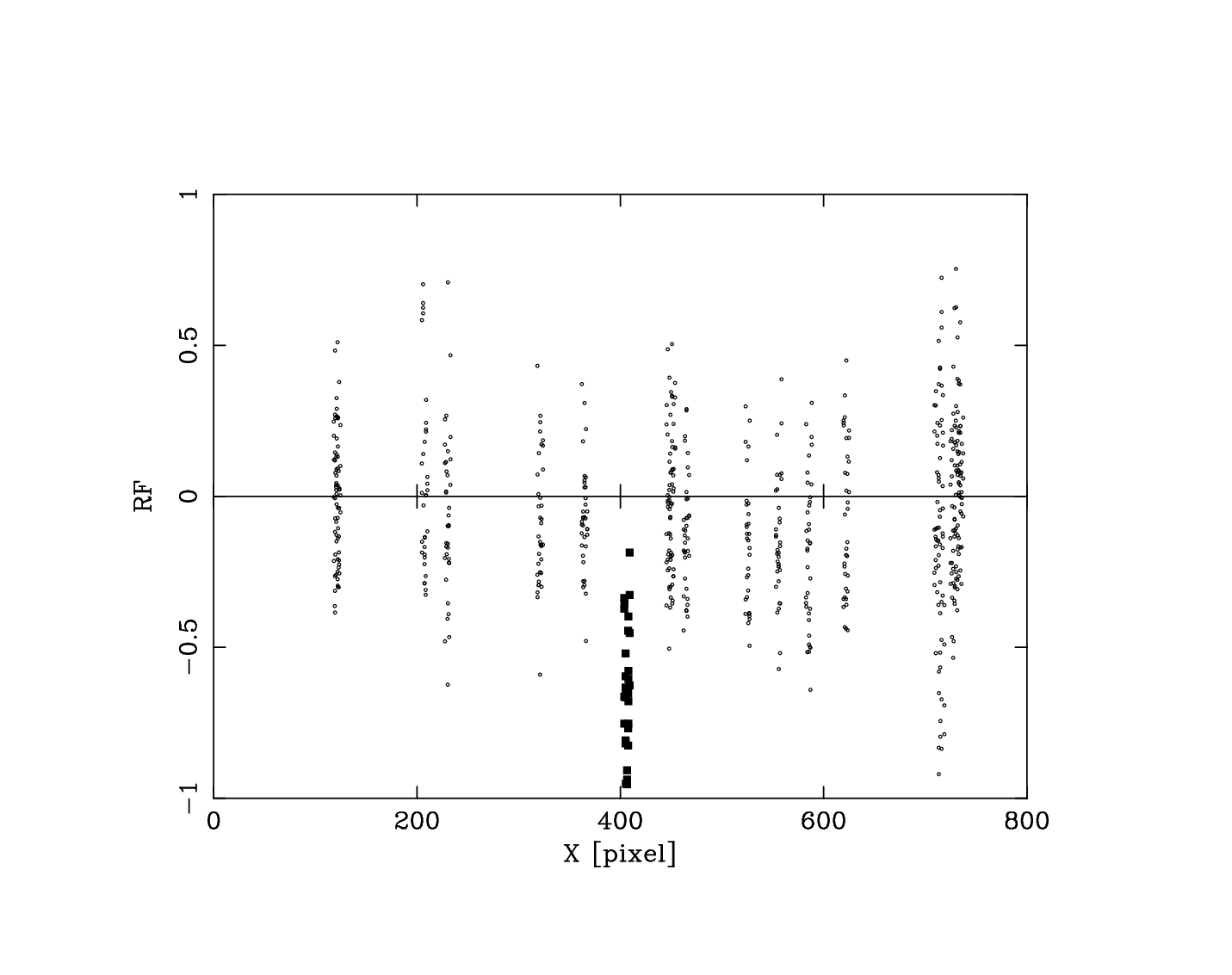}
}
\subfigure[]
{\label{pc-10-rf-y}
\includegraphics[angle=-90,scale=0.28]{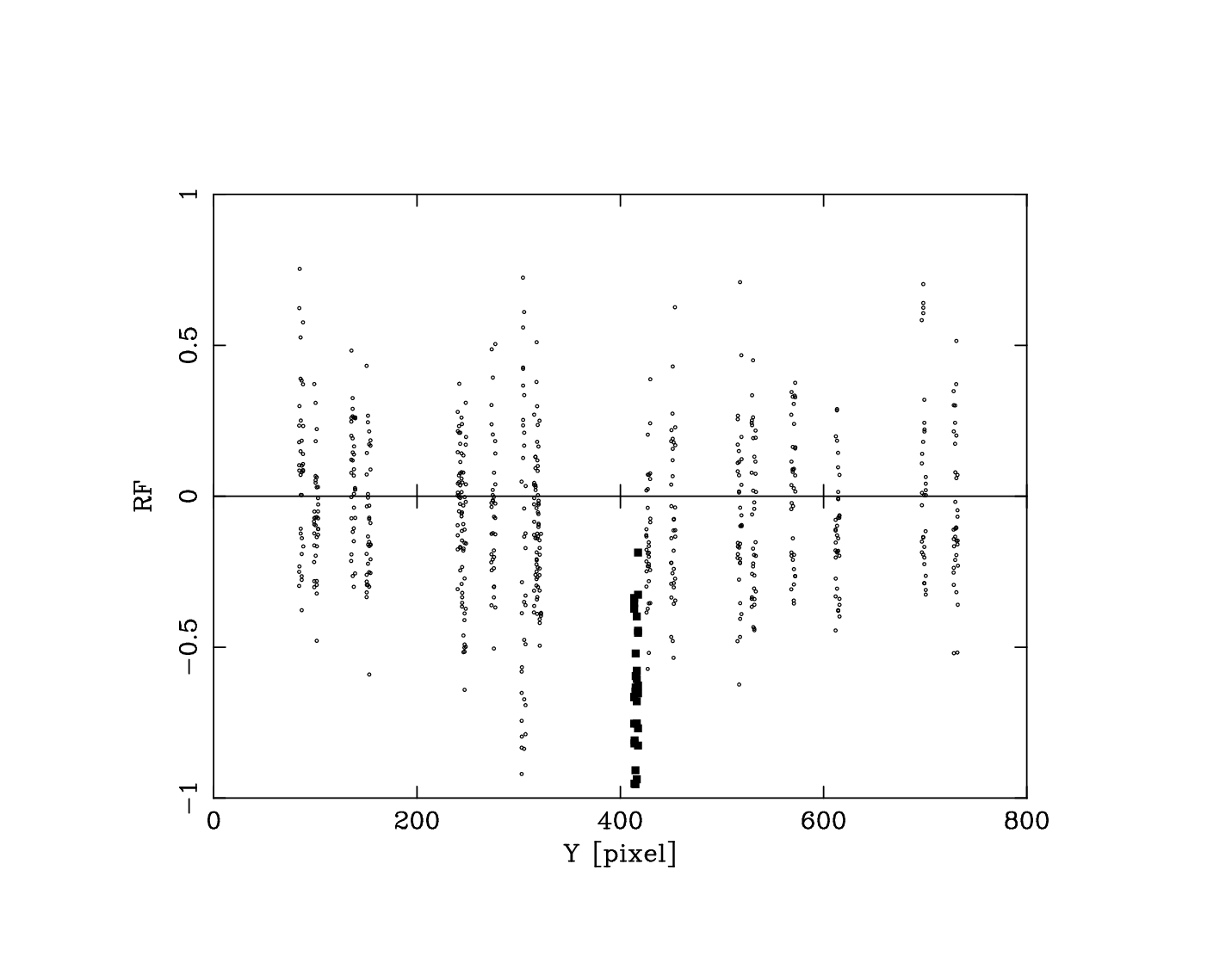}
}
\caption{Flux residual \textit{versus} location for objects in an
image of the UMI~J$1508+6717$ field.  A plot in the top row displays
points from 40 images and a plot in the remaining rows displays points
from 36 images.  The left panels plot ${\cal RF}$ \textit{versus} the
$X$-coordinate and the right panels plot ${\cal RF}$ \textit{versus}
the $Y$-coordinate.  The solid square represents the QSO.  Panels a)
and b) correspond to the 1999 epoch; c) and d) to the 2001 epoch; and
e) and f) to the 2003 epoch.  For ease of comparison, all of the plots
have the same scale on the vertical axis.}
\label{rf-pc}
\end{figure}
\clearpage
\begin{figure}
\subfigure[]
{\label{stis-8-rxry}
\includegraphics[angle=-90,scale=0.4]{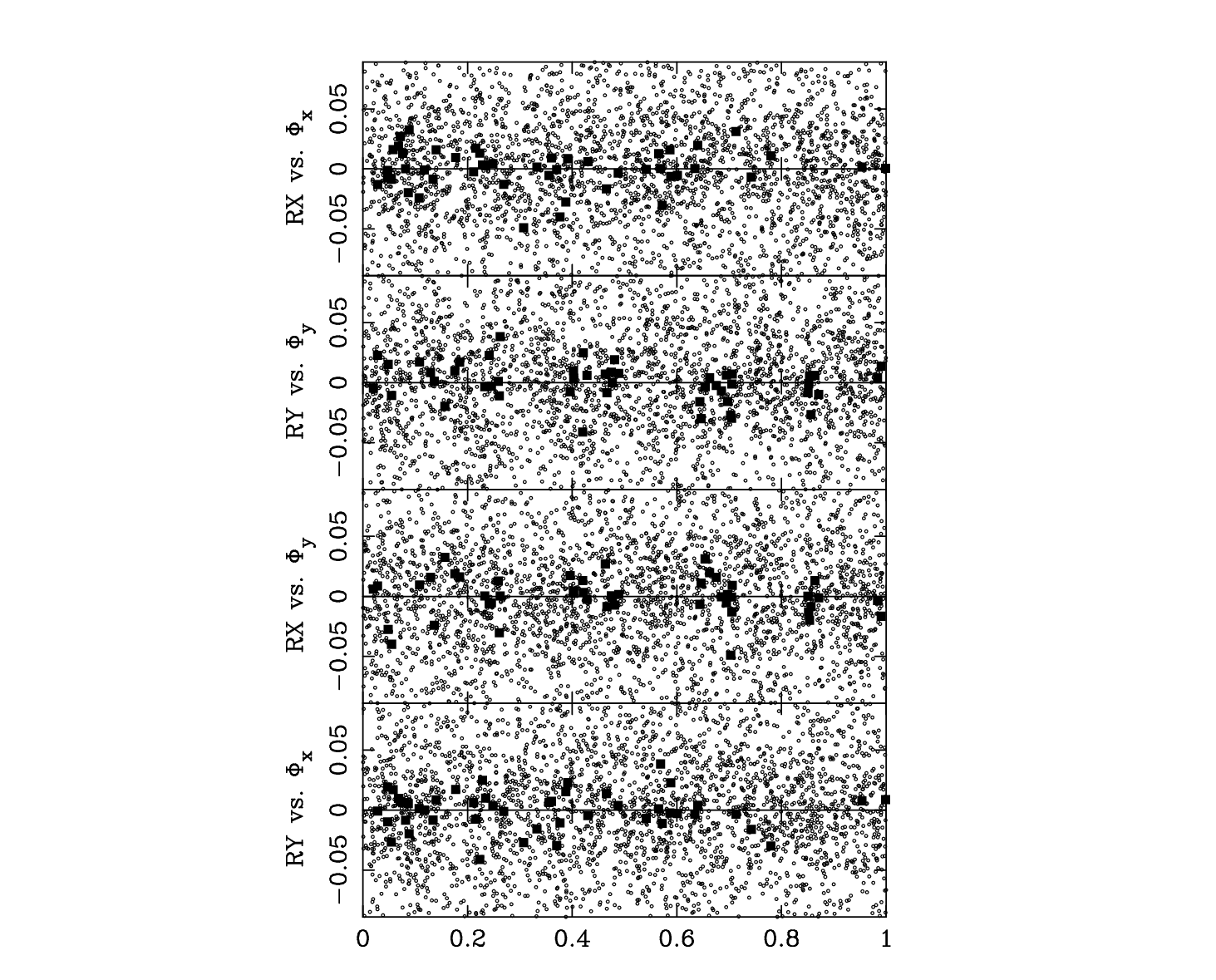}
}
\subfigure[]
{\label{stis-9-rxry}
\includegraphics[angle=-90,scale=0.4]{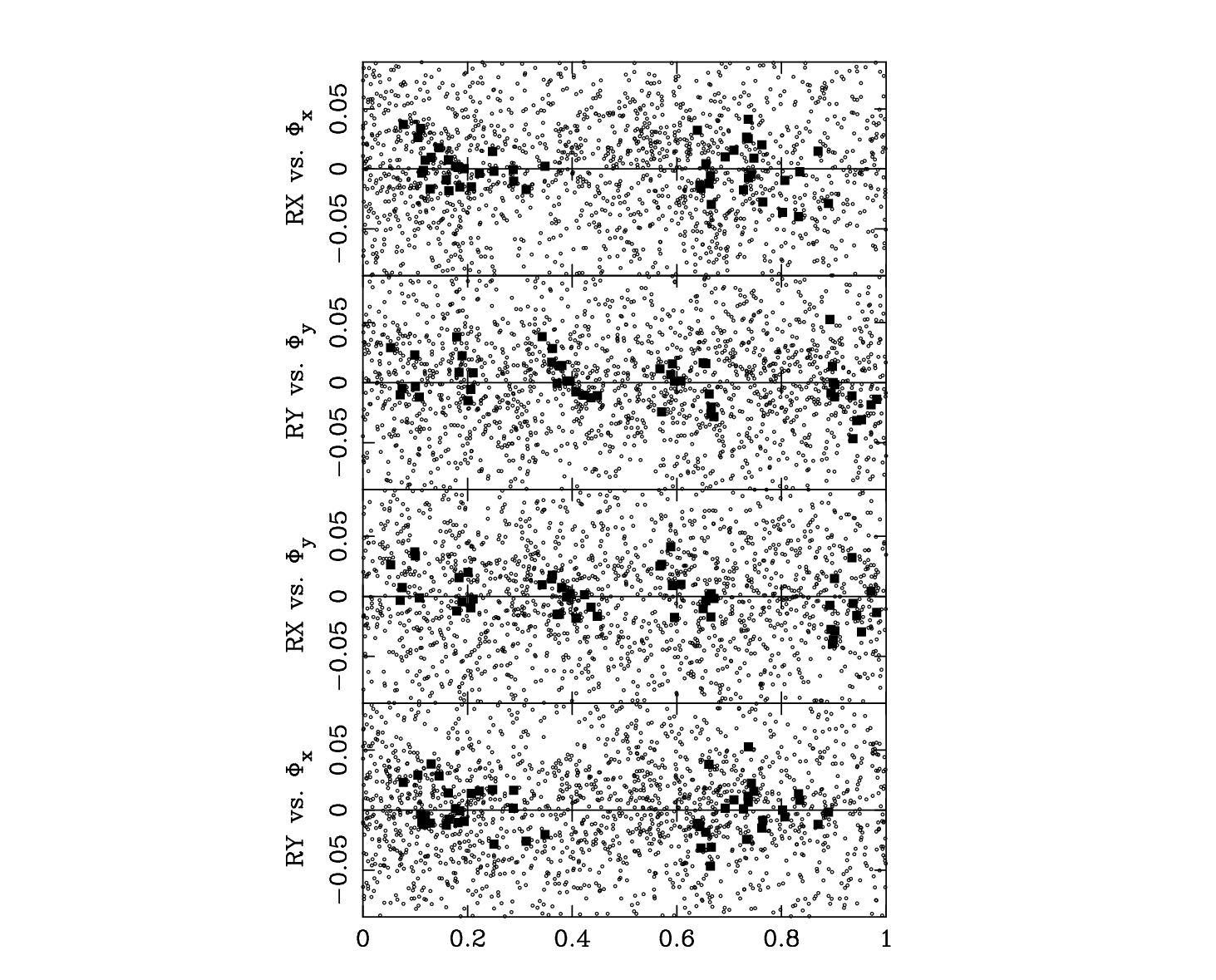}
}
\subfigure[]
{\label{stis-10-rxry}
\includegraphics[angle=-90,scale=0.4]{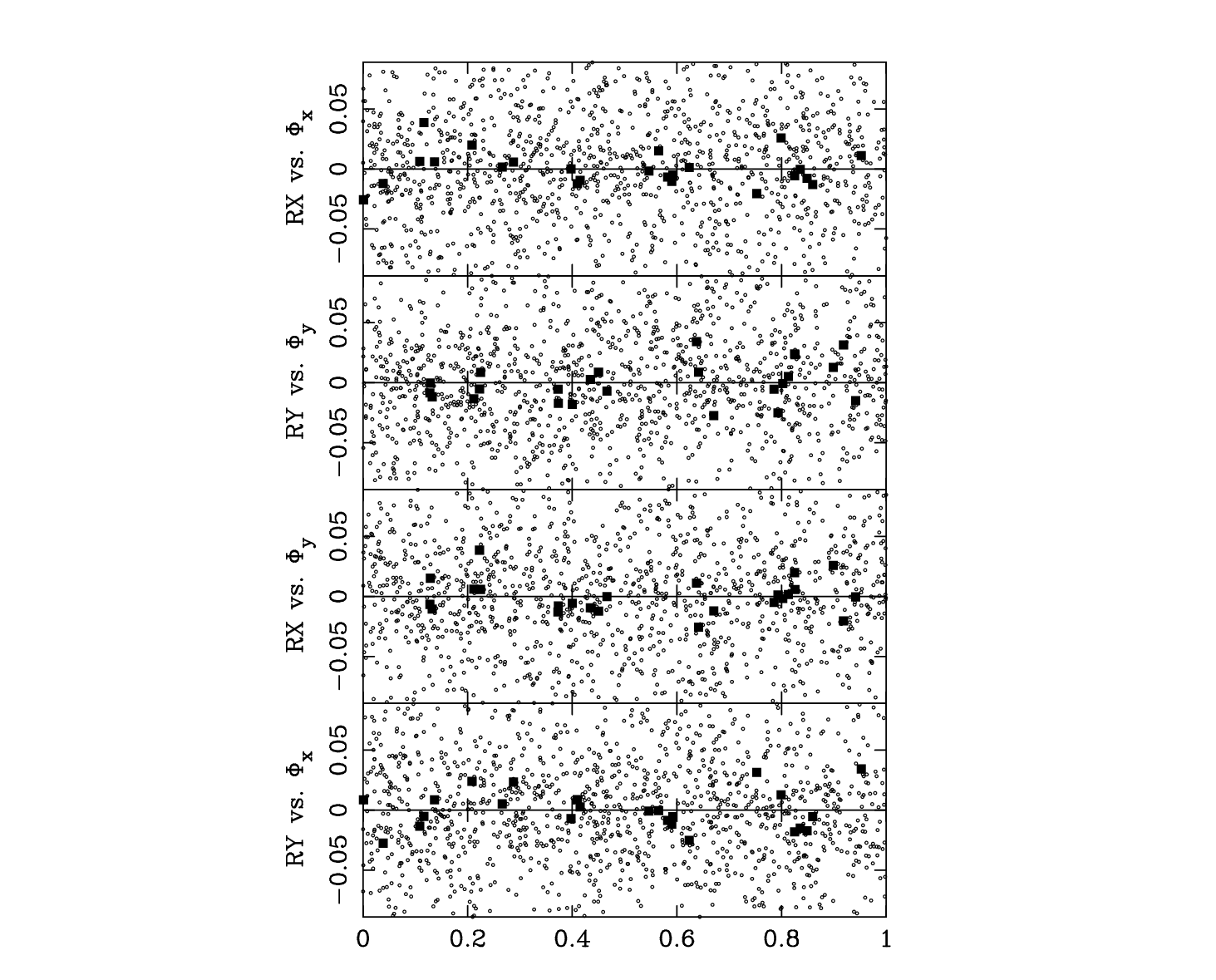}
}
\caption{Plots for the UMI~J$1508+6716$ field of the position
residuals, ${\cal RX}$ and ${\cal RY}$, as a function of the pixel
phase, $\Phi_{x}$ and $\Phi_{y}$.  The panels a), b), and c)
correspond to the epochs 2000, 2001, and 2002, respectively.  The
solid square corresponds to the QSO.}
\label{stis-rxry}
\end{figure}
\clearpage
\begin{figure}
\subfigure[]
{\label{pc-7-rxry}
\includegraphics[angle=-90,scale=0.4]{f7a.eps}
}
\subfigure[]
{\label{pc-9-rxry}
\includegraphics[angle=-90,scale=0.4]{f7b.eps}
}
\subfigure[]
{\label{pc-10-rxry}
\includegraphics[angle=-90,scale=0.4]{f7c.eps}
}
\caption{Plots for the UMI~J$1508+6717$ field of the position
residuals, ${\cal RX}$ and ${\cal RY}$, as a function of the pixel
phase, $\Phi_{x}$ and $\Phi_{y}$.  The panels a), b), and c)
correspond to the epochs 1999, 2001, and 2003, respectively.  The
solid square corresponds to the QSO.}
\label{pc-rxry}
\end{figure}
\clearpage
\begin{figure}
\subfigure[]
{\label{stis-8-rms-sn}
\includegraphics[angle=-90,scale=0.4]{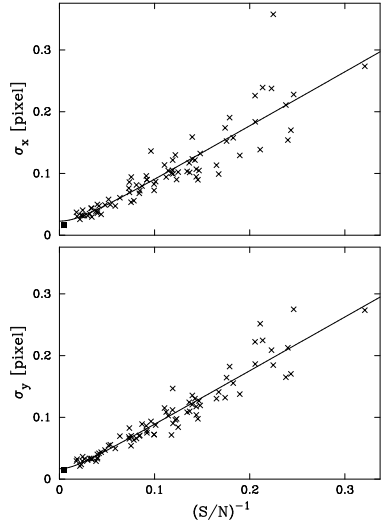}
}
\subfigure[]
{\label{stis-9-rms-sn}
\includegraphics[angle=-90,scale=0.4]{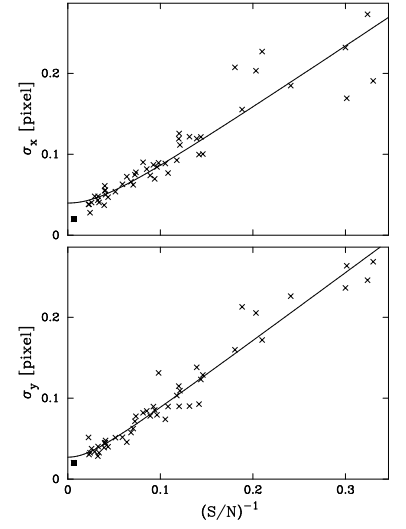}
}
\subfigure[]
{\label{stis-10-rms-sn}
\includegraphics[angle=-90,scale=0.4]{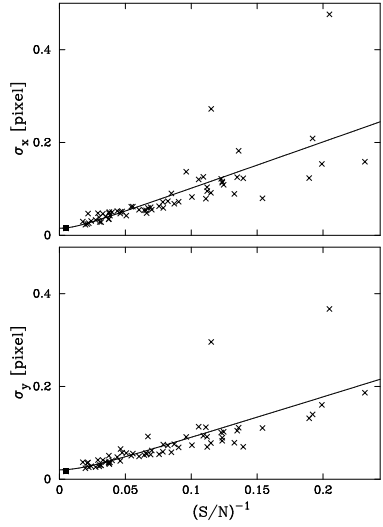}
}
\caption{Plots the $rms$ scatter around the mean of the $X$-component
(top panel) and the $Y$-component (bottom panel) of the centroid as a
function of $(S/N)^{-1}$ for the UMI~J$1508+6716$ field.  The solid
square corresponds to the QSO.  (a) For the epoch 2000.  (b) For the
epoch 2001.  (c) For the epoch 2002.}
\label{stis-rms-sn}
\end{figure}
\clearpage
\begin{figure}
\subfigure[]
{\label{pc-7-rms-sn}
\includegraphics[angle=-90,scale=0.4]{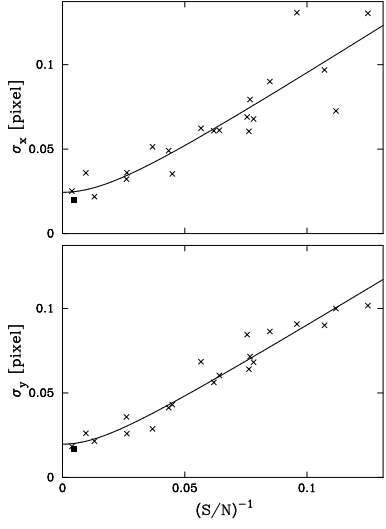}
}
\subfigure[]
{\label{pc-9-rms-sn}
\includegraphics[angle=-90,scale=0.4]{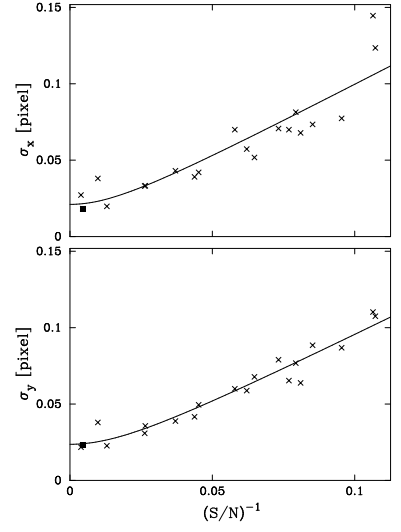}
}
\subfigure[]
{\label{pc-10-rms-sn}
\includegraphics[angle=-90,scale=0.4]{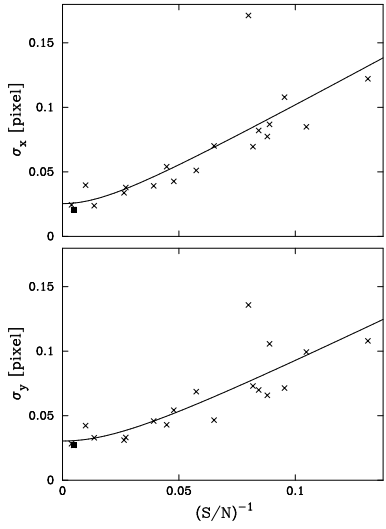}
}
\caption{Same as Figure~\ref{stis-rms-sn} for the UMI~J$1508+6717$
field.  (a) For the epoch 1999.  (b) For the epoch 2001.  (c) For the
epoch 2003.}
\label{pc-rms-sn}
\end{figure}
\clearpage
\begin{figure}
\includegraphics[angle=-90,scale=0.8]{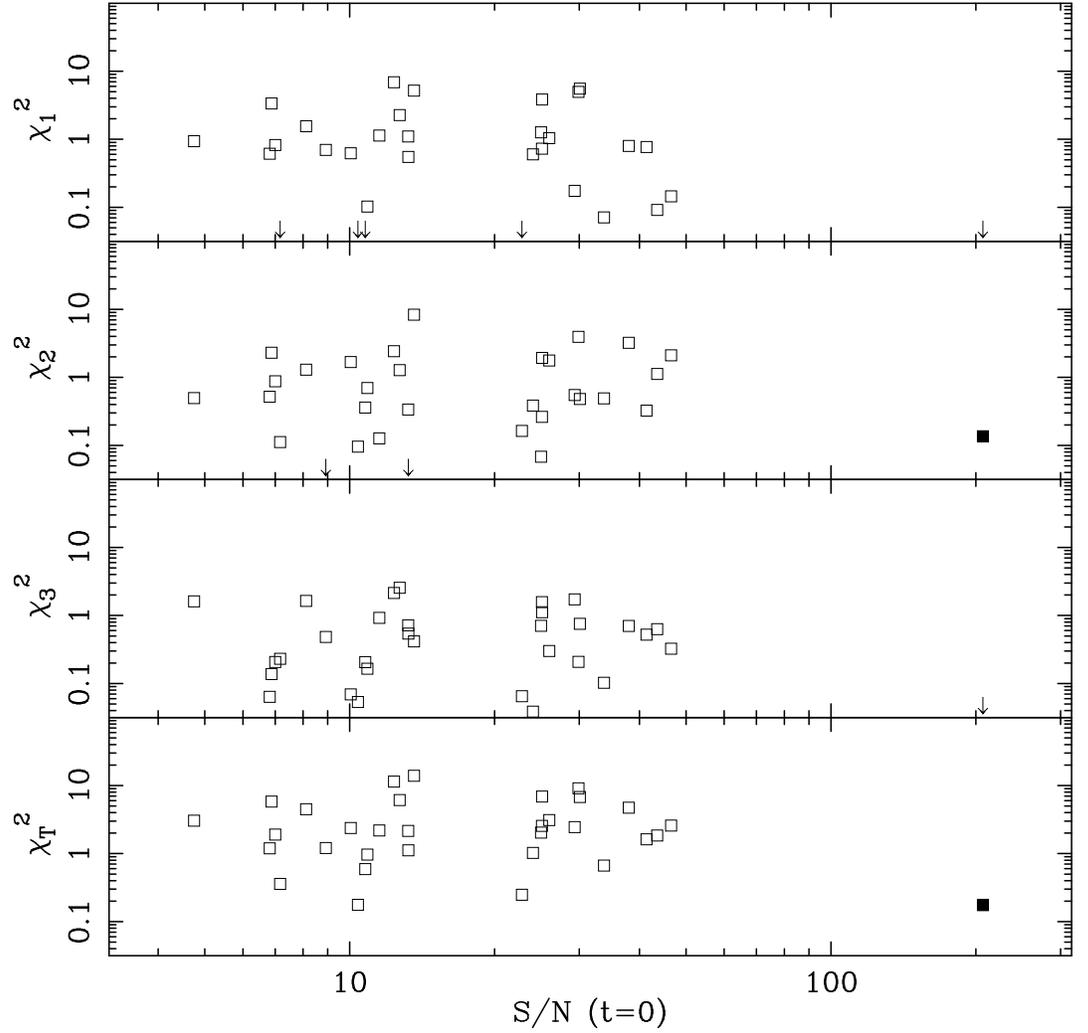}
\caption{The $\chi^{2}$ of an object in the UMI~J$1508+6716$ field,
defined by equation~\ref{eq:k2}, \textit{versus} its $S/N$ at the
first epoch $(t=0)$.  The top three panels show the contributions to
the $\chi^2$ from centroids measured at epochs $j=1,2,3$,
respectively, and the bottom panel is the total $\chi^{2}$.  An arrow
pointing down indicates a point outside of the plot range.  The solid
square represents the QSO.  The gap in the distribution of points
between a $S/N$ of about 10 and about 20 is an artefact of the data.}
\label{fig:k2stis}
\end{figure}
\clearpage
\begin{figure}
\includegraphics[angle=-90,scale=0.8]{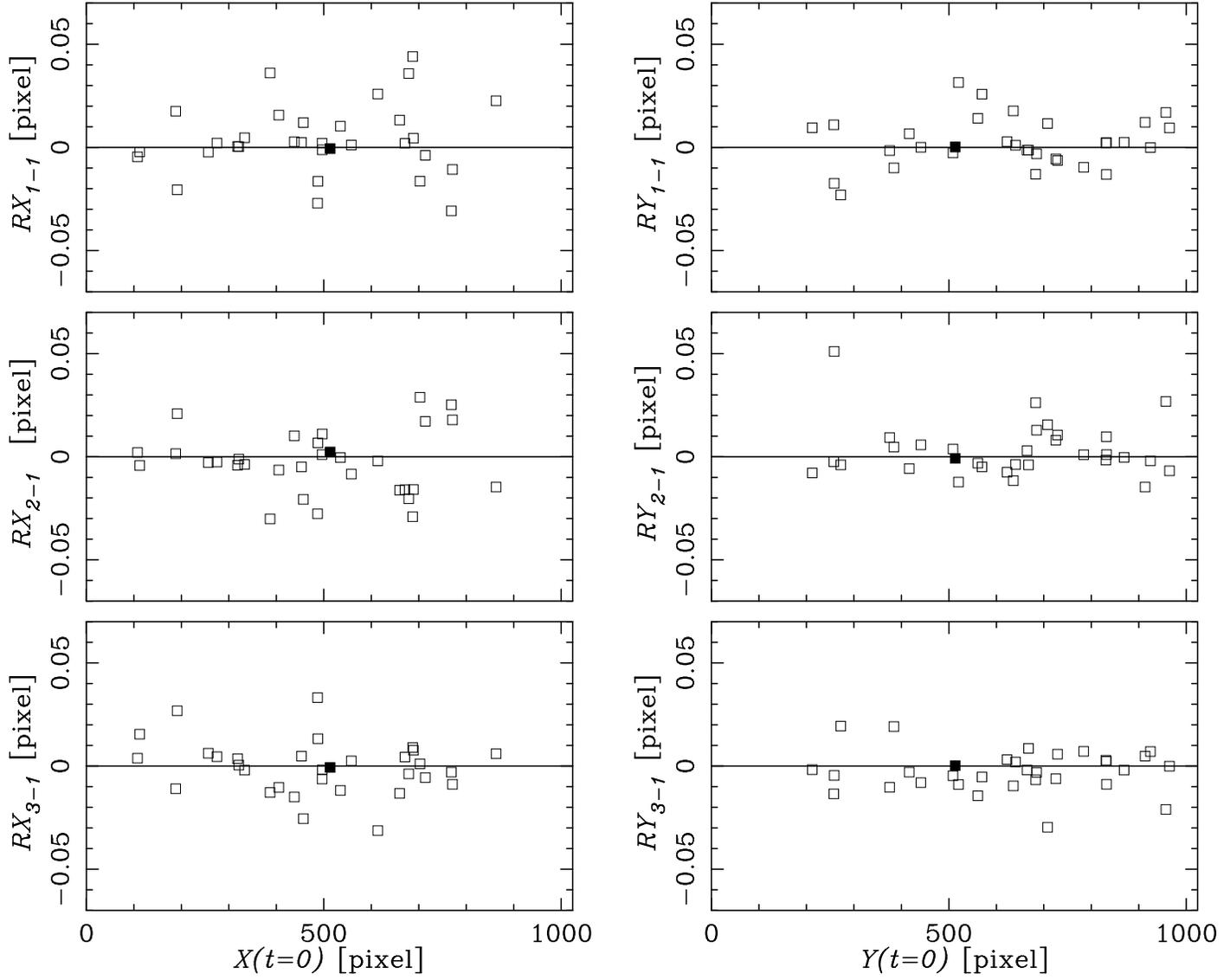}
\caption{Position residuals defined by the Equations~\ref{eq:rx} and
\ref{eq:ry} for the objects in the UMI~J$1508+6716$ field.  From top
to bottom, the panels are for the first, second, and third epoch,
respectively.  The panels on the left show $RX$ \textit{versus} $x$
and those on the right show $RY$ \textit{versus} $y$.  The solid
square corresponds to the QSO.}
\label{fig:stis-rxry}
\end{figure}
\clearpage
\begin{figure}
\centering
\includegraphics[angle=-90,scale=0.8]{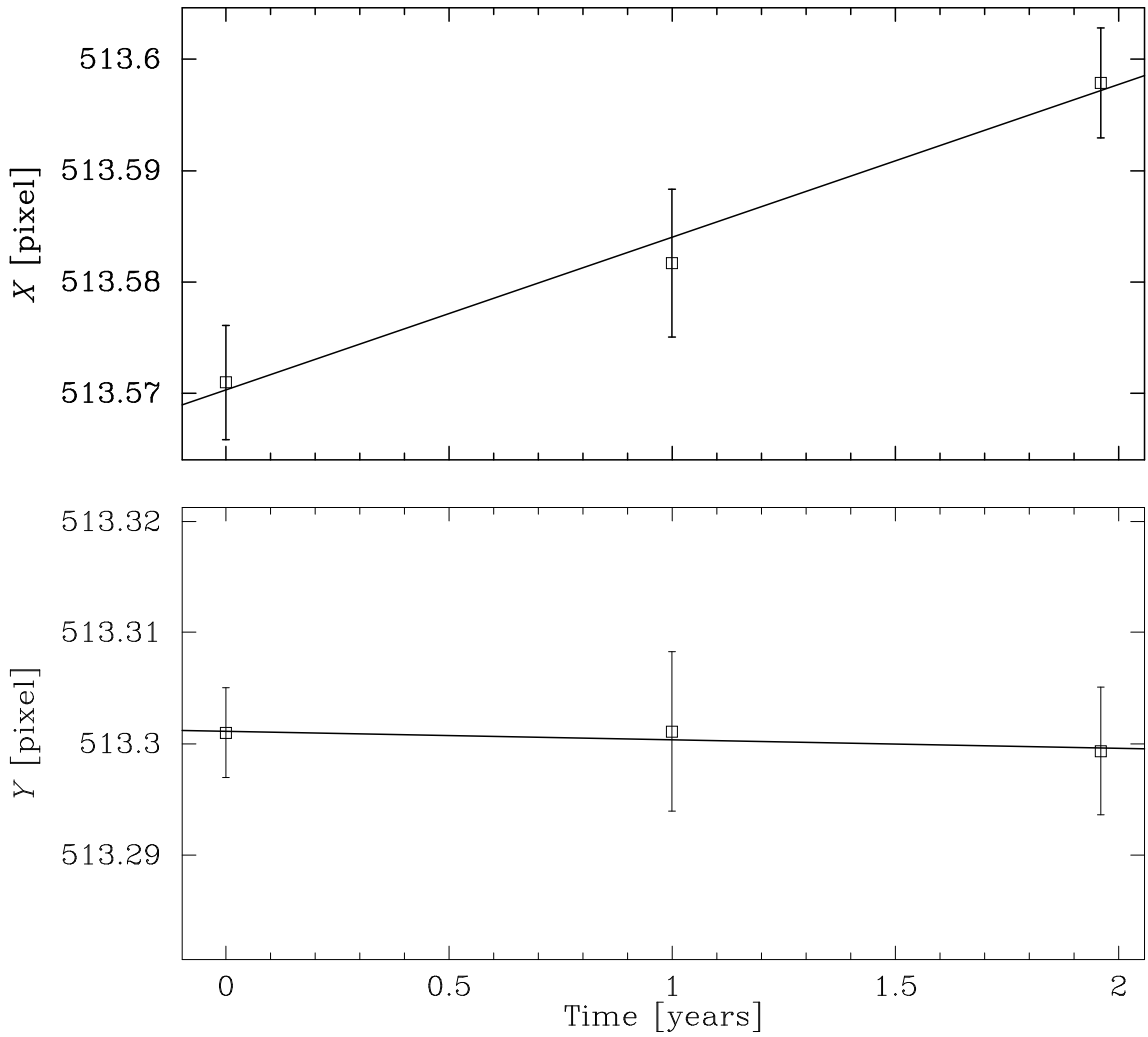}
\caption{The location of the QSO as a function of time for the
UMI~J$1508+6716$ field in the standard coordinate system.  The
vertical axis in each plot has the same scale.}
\label{xqyqstis}
\end{figure}
\clearpage
\begin{figure}
\centering
\includegraphics[angle=-90,scale=0.8]{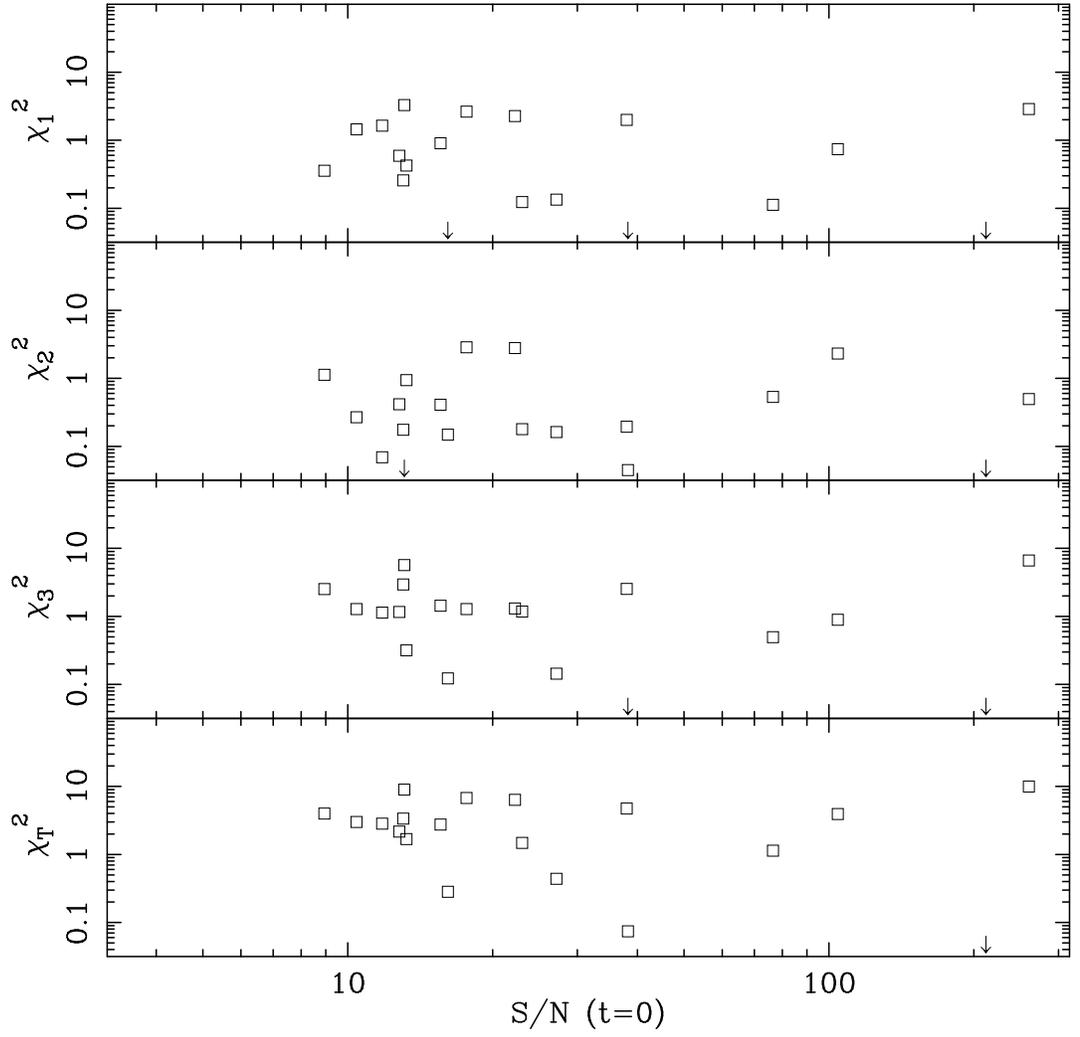}
\caption{The figure for the UMI~J$1508+6717$ field that is analogous
to Figure~\ref{fig:k2stis}.  Note that the point corresponding to the
QSO is below the lower limit of the plots, as indicated by the arrow
at a S/N of 214.}
\label{fig:k2pc}
\end{figure}
\clearpage
\begin{figure}
\centering
\includegraphics[angle=-90,scale=0.8]{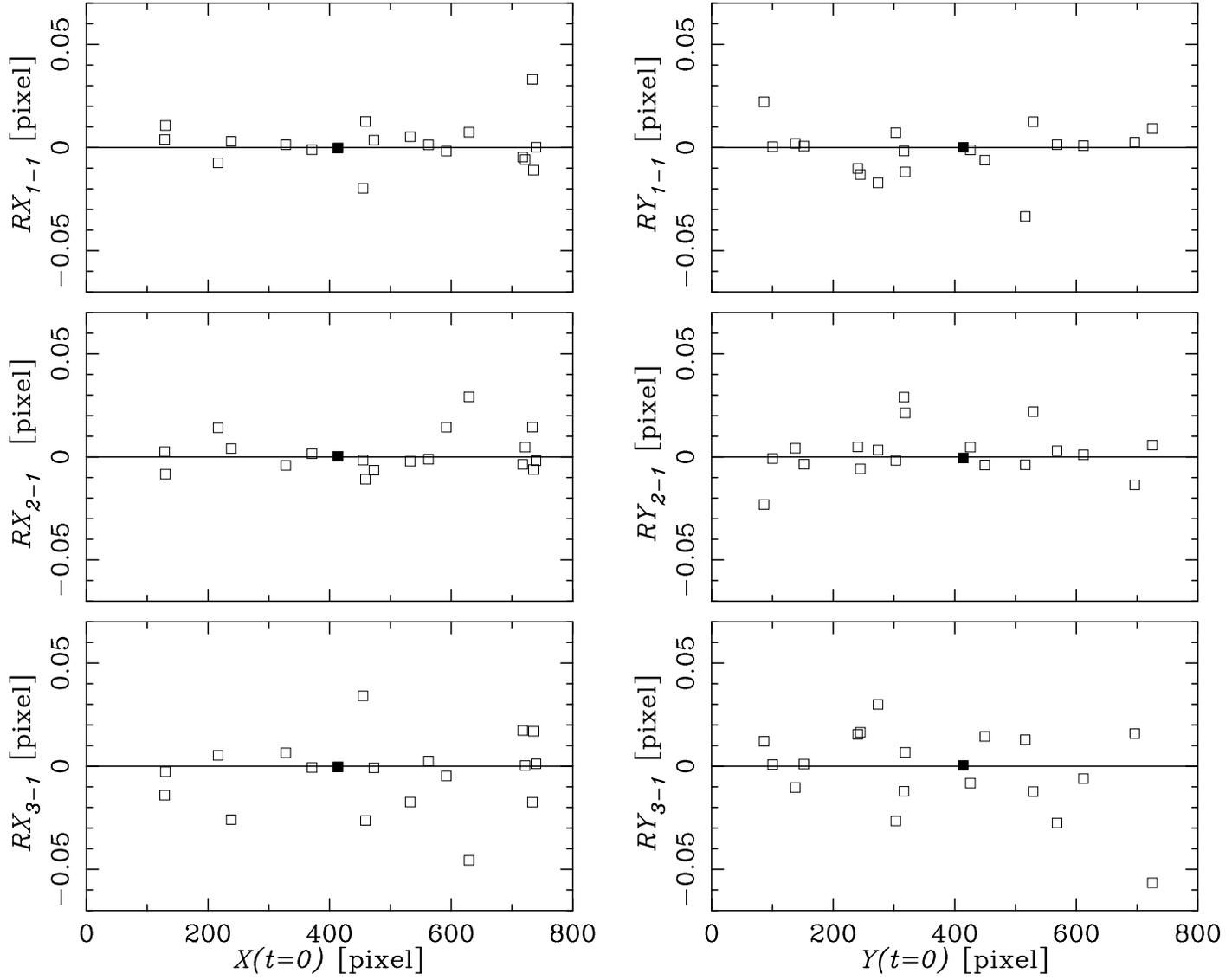}
\caption{Plots of the position residuals for the UMI~J$1508+6717$
field analogous to Figure~\ref{fig:stis-rxry}.}
\label{fig:pc-rxry}
\end{figure}
\clearpage
\begin{figure}
\centering
\includegraphics[angle=-90,scale=0.8]{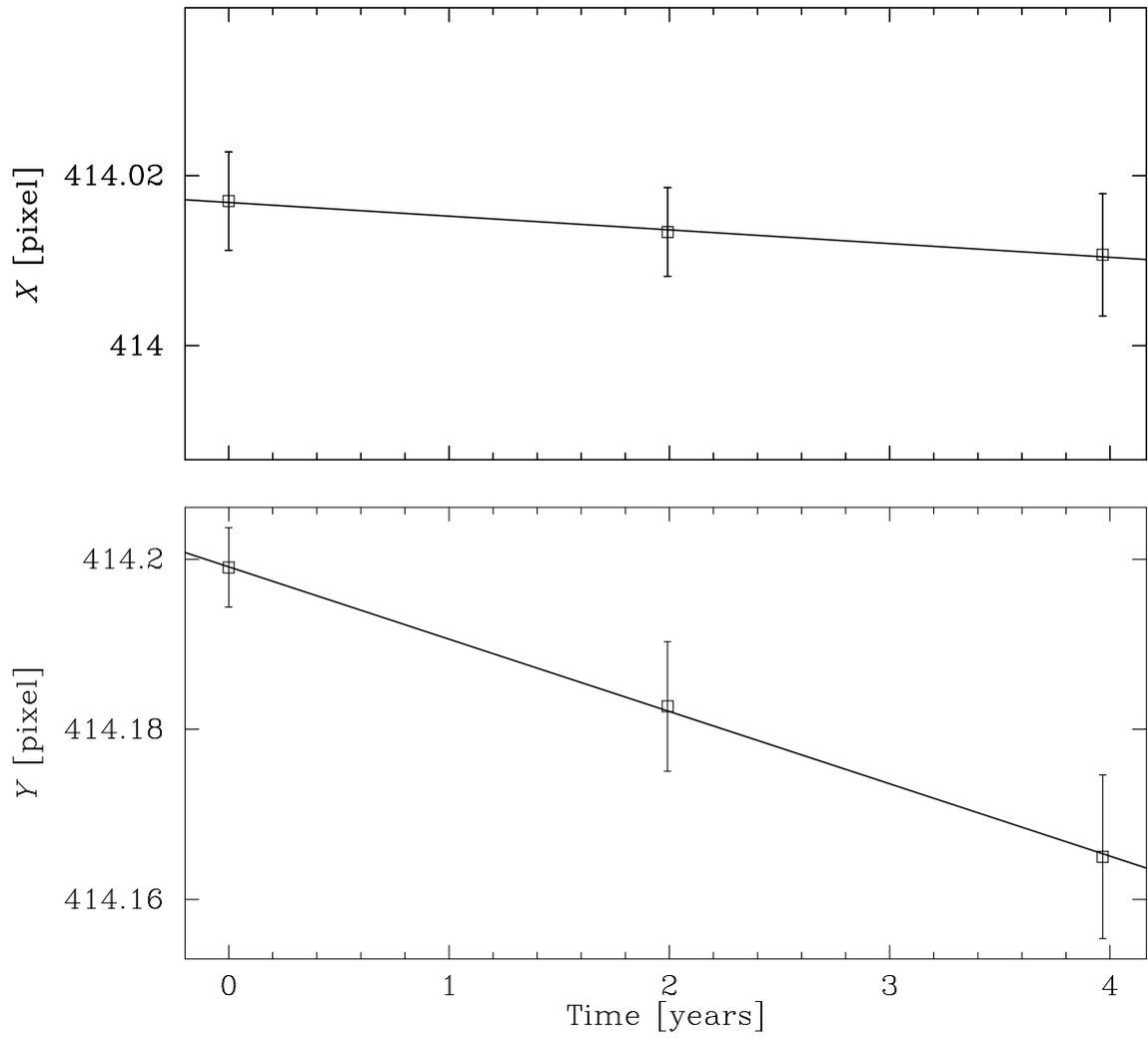}
\caption{An analogous plot to Figure~\ref{xqyqstis}\ for the
UMI~J$1508-6717$ field.}
\label{xqyqpc}
\end{figure}
\clearpage
\begin{figure}
\centering
\includegraphics[angle=-90,scale=0.8]{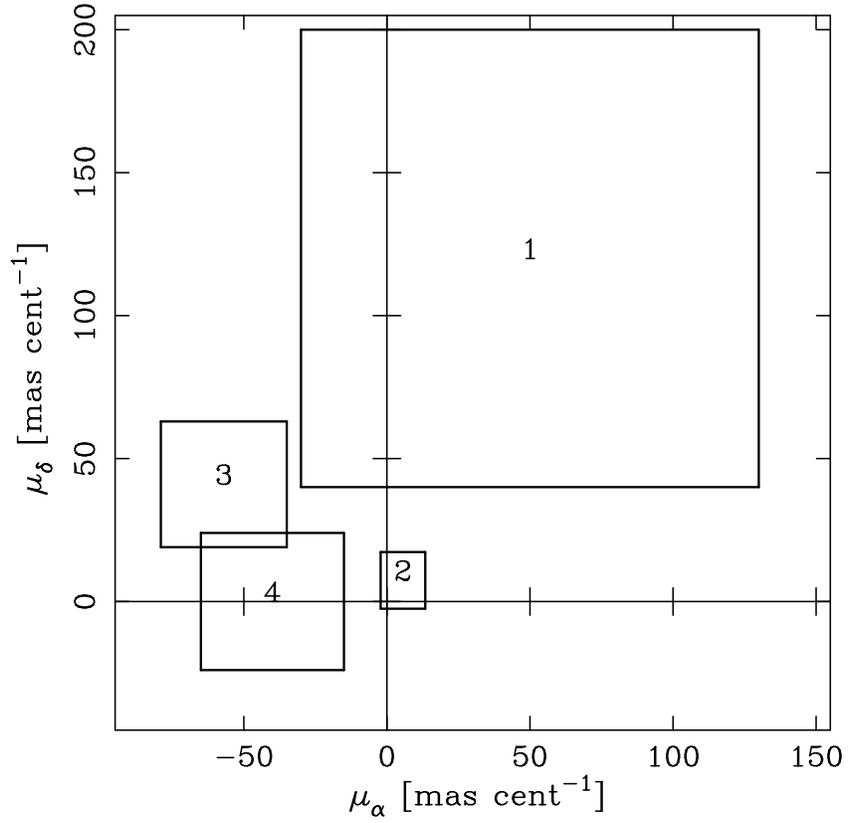}
\caption{Comparison of four independent measurements of the proper
motion of Ursa Minor.  The center of a rectangle is the best estimate
of the proper motion and the sides are offset by the 1-$\sigma$
uncertainties.  Rectangles 1, 2, 3, and 4 correspond to the
measurements by Scholz \& Irwin (1993), Schweitzer (1996), this study
(field UMI~J$1508+6716$), and this study (UMI~J$1508+6717$).}
\label{mus-comp}
\end{figure}
\clearpage
\begin{figure}
\centering
\includegraphics[angle=-90,scale=0.8]{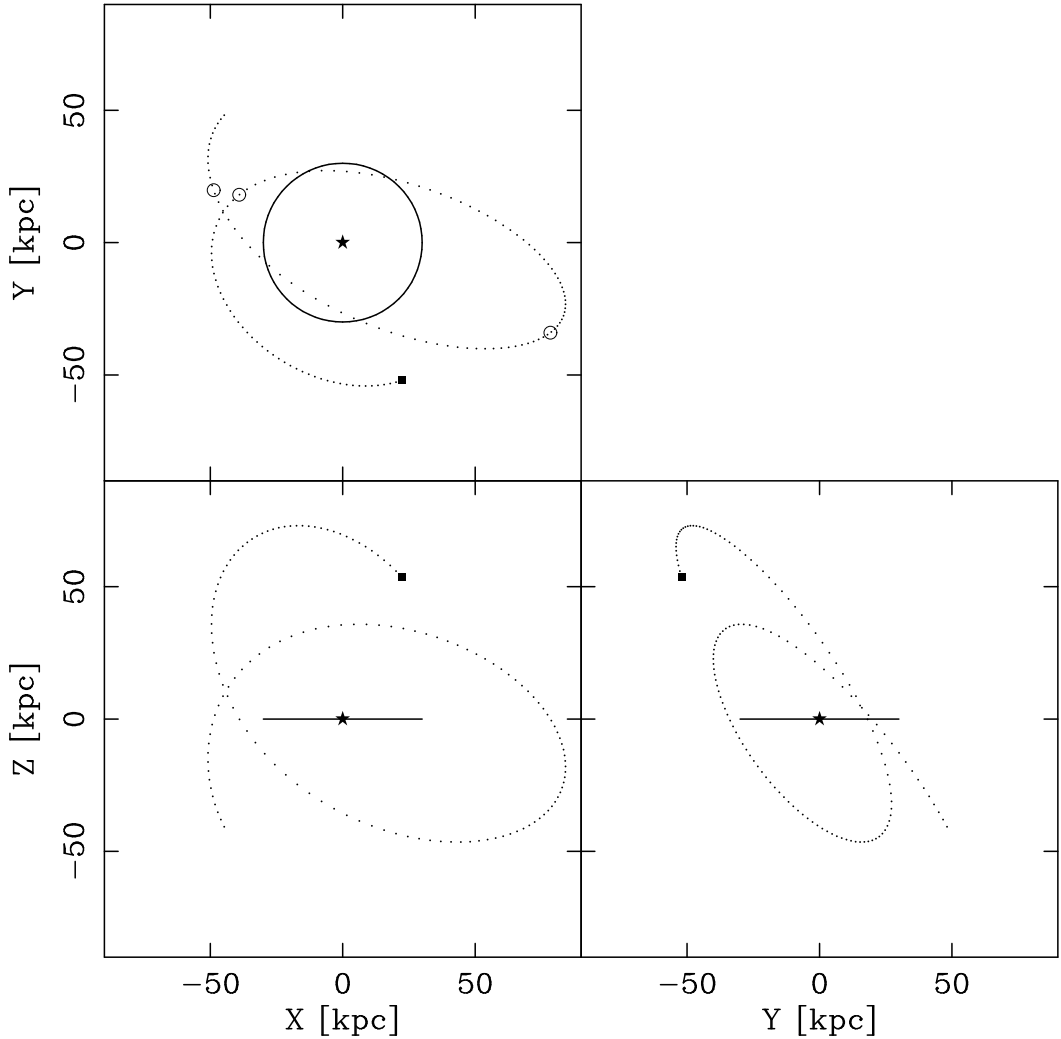}
\caption{Projections of the orbit of Ursa Minor onto the $X-Y$ plane
(top-left panel), the $X-Z$ plane (bottom-left panel), and the $Y-Z$
plane (bottom-right panel).  The origin of the right-handed coordinate
system is at the Galactic center, which is marked with a solid star.
The Galactic disk is in the $X-Y$ plane and the present location of
the Sun is on the positive $X$ axis.  The solid square marks the
current location of Ursa Minor at $(X,Y,Z)=(22,-52,54)$~kpc.  For
reference, the large circle in the $X-Y$ plane has a radius of 30~kpc.
The three small circles in the $X-Y$ projection mark the points where
Ursa Minor passes through the plane of the Galactic disk.  The
integration starts from the present and extends backwards in time for
3~Gyr.}
\label{orbit}
\end{figure}
\clearpage

\setcounter{table}{0}
\newdimen\digitwidth\setbox0=\hbox{\rm 0}\digitwidth=\wd0
\catcode`@=\active\def@{\kern\digitwidth}

\begin{deluxetable}{lccccc}
\tablecolumns{6}
\tablewidth{5.5truein} 
\tablecaption{Fitted free parameters}
\tablehead{ & & \multicolumn{2}{c}{$X$}&\multicolumn{2}{c}{$Y$}\\
\cline{3-4} \cline{5-6} & & \colhead{$a$}&\colhead{$\sigma_0$}&\colhead{$a$}
&\colhead{$\sigma_0$}\\
\colhead{Field}&\colhead{Epoch}&\colhead{pixel}&\colhead{pixel}&
\colhead{pixel}&\colhead{pixel}\\
\colhead{(1)}&\colhead{(2)}&\colhead{(3)}&\colhead{(4)}&\colhead{(5)}&
\colhead{(6)}}
\startdata
                  & 2000 & 0.879 & 0.023 & 0.874 & 0.017 \\
UMI@J$1508+6716$  & 2001 & 0.768 & 0.040 & 0.846 & 0.027 \\
                  & 2002 & 1.004 & 0.015 & 0.882 & 0.020 \\
\noalign{\vspace{1.5pt}}
\hline
\noalign{\vspace{1.5pt}}
                  & 1999 & 0.921 & 0.024 & 0.881 & 0.020 \\
UMI@J$1508+6717$  & 2001 & 0.974 & 0.021 & 0.926 & 0.024 \\
                  & 2003 & 0.987 & 0.025 & 0.878 & 0.030 \\
\enddata
\end{deluxetable}

\begin{deluxetable}{lrr}
\tablecolumns{3}
\tablewidth{3.5truein} 
\tablecaption{Measured Proper Motion of Ursa Minor}
\tablehead{
&\colhead{$\mu_{\alpha}$}&\colhead{$\mu_{\delta}$}\\ 
\colhead{Field}&\multicolumn{2}{c}{(mas cent$^{-1}$)}\\
\colhead{(1)}&\colhead{(2)}&\colhead{(3)}}
\startdata
UMI@J$1508+6716$  &$-57\pm22$&$41\pm22$\\
\noalign{\vspace{3pt}}
UMI@J$1508+6717$ &$-40\pm25$&$ 0\pm24$ \\
\noalign{\vspace{3pt}}
\hline
\noalign{\vspace{5pt}}
Weighted Mean: &$-50\pm 17$ & $22\pm 16$ \\
\enddata
\end{deluxetable}
\clearpage

\begin{deluxetable}{ccccccc}
\tablecolumns{7}
\tablewidth{5.0truein} 
\tablecaption{$\!\!\rm{A}$~ Measured Proper Motions For Objects in
the UMI~J$1508+6716$ Field}
\tablehead{ &X&Y& &$\mu_{\alpha}$&$\mu_{\delta}$ &  \\
\colhead{ID}&\colhead{(pixels)}&\colhead{(pixels)}&\colhead{$S/N$}&\colhead{(mas
@cent$^{-1}$)}&
\colhead{(mas@cent$^{-1}$)} & \colhead{$\chi^2$} \\
\colhead{(1)}&\colhead{(2)}&\colhead{(3)}&\colhead{(4)}&\colhead{(5)}&
\colhead{(6)} & \colhead{(7)}
} 
\startdata
  1& 514& 513& 207& $    0 \pm  28 $& $    0 \pm  27 $&  0.17 \\
  2& 671& 640&  47& $   59 \pm  35 $& $ -140 \pm  34 $&  2.58 \\
  3& 321& 375&  44& $   -2 \pm  39 $& $  161 \pm  39 $&  1.86 \\
  4& 771& 685&  38& $-1511 \pm  48 $& $-1573 \pm  46 $&  4.72 \\
  5& 497& 831&  34& $  -88 \pm  41 $& $ -658 \pm  41 $&  0.66 \\
  6& 318& 665&  23& $-1027 \pm  48 $& $   61 \pm  48 $&  0.25 \\
  7& 689& 623&  11& $   67 \pm  69 $& $ -216 \pm  66 $&  0.96 \\
  8& 453& 509&  10& $ -619 \pm  94 $& $ -649 \pm  95 $&  1.27 \\
\enddata
\end{deluxetable}
\setcounter{table}{2}
\begin{deluxetable}{ccccccc}
\tablecolumns{7}
\tablewidth{5.0truein} 
\tablecaption{$\!\!\rm{B}$~ Measured Proper Motions For Objects in
the UMI~J$1508+6717$ Field}
\tablehead{ &X&Y& &$\mu_{\alpha}$&$\mu_{\delta}$ &  \\
\colhead{ID}&\colhead{(pixels)}&\colhead{(pixels)}&\colhead{$S/N$}&\colhead{(mas
@cent$^{-1}$)}&
\colhead{(mas@cent$^{-1}$)} & \colhead{$\chi^2$} \\
\colhead{(1)}&\colhead{(2)}&\colhead{(3)}&\colhead{(4)}&\colhead{(5)}&
\colhead{(6)} & \colhead{(7)}
}
\startdata
  1& 414& 414& 212& $    0 \pm  21 $& $    0 \pm  20 $&  0.00 \\
  2& 563& 426&  77& $ -166 \pm  23 $& $ -441 \pm  20 $&  2.46 \\
\enddata
\end{deluxetable}

\hoffset=-0.5truein
\begin{deluxetable}{lrrrrrrrrr}
\tablecolumns{10}
\tablewidth{8.5truein} 
\tablecaption{Galactic-Rest-Frame Proper Motion and Space Velocity of
Ursa Minor}
\tablehead{&\colhead{$\mu_{\alpha}^{\mbox{\tiny{Grf}}}$}&
\colhead{$\mu_{\delta}^{\mbox{\tiny{Grf}}}$}&
\colhead{$\mu_{l}^{\mbox{\tiny{Grf}}}$}&\colhead{$\mu_{b}^{\mbox{\tiny{Grf}}}$}&
\colhead{$\Pi$}&\colhead{$\Theta$}&\colhead{$Z$}&\colhead{$V_{r}$}&
\colhead{$V_{t}$}\\ 
\colhead{Field}&\multicolumn{2}{c}{(mas cent$^{-1}$)}
&\multicolumn{2}{c}{(mas cent$^{-1}$)}&\colhead{(km s$^{-1}$)}&\colhead{(km s$^{-1}$)}&\colhead{(km s$^{-1}$)}
&\colhead{(km s$^{-1}$)}&\colhead{(km s$^{-1}$)}\\
\colhead{(1)}&\colhead{(2)}&\colhead{(3)}&\colhead{(4)}
&\colhead{(5)}&\colhead{(6)}&\colhead{(7)}&\colhead{(8)}&\colhead{(9)}
&\colhead{(10)}}
\startdata
UMI@J$1508+6716$&$-15\pm22$&$56\pm22$&$50\pm22$&$-29\pm22$&$36\pm61$&$-176\pm74$&$-138\pm56$&$-68\pm58$&$216\pm69$\\
\noalign{\vspace{5pt}}
UMI@J$1508+6717$&$2\pm25$&$16\pm24$&$9\pm25$&$-13\pm25$&$-26\pm69$&$-38\pm84$&$-96\pm63$&$-85\pm66$&$64\pm73$\\
\noalign{\vspace{3pt}}
\hline
\noalign{\vspace{5pt}}
Weighted Mean:&$-8\pm17$&$38\pm16$&$32\pm17$&$-22\pm17$&$9\pm46$&$-116\pm56$&$-119\pm42$&$-75\pm44$&$144\pm50$\\
\enddata
\end{deluxetable}

\begin{deluxetable}{lcccc}
\tablecolumns{5}
\tablewidth{4.75truein} 
\tablecaption{Orbital elements of Ursa Minor}
\tablehead{Quantity&Symbol&Unit&Value&95\% Conf. Interv.\\
\colhead{(1)}&\colhead{(2)}&\colhead{(3)}&\colhead{(4)}&\colhead{(5)}}
\startdata
Perigalacticon&$R_{p}$&kpc&$40$&$(10,76)$ \\
\noalign{\vspace{1pt}}
Apogalacticon&$R_{a}$&kpc&$89$&$(78,160)$\\
\noalign{\vspace{1pt}}
Eccentricity &$e$&&$0.39$&$(0.09,0.79)$\\
\noalign{\vspace{1pt}}
Period&$T$&Gyr&$1.5$&$(1.1,2.7)$\\
\noalign{\vspace{1pt}}
Inclination&$\Phi$&deg&124&$(94,136)$ \\
\noalign{\vspace{1pt}}
Longitude&$\Omega$&deg& 153&$(116,193)$ \-
\enddata
\end{deluxetable}

\end{document}